\documentclass[aps,prl,showpacs,twocolumn,superscriptaddress]{revtex4-1}
\usepackage{amsmath,amssymb,amsfonts,bm}
\usepackage{graphicx}
\usepackage{epstopdf}
\usepackage{dcolumn}
\usepackage{mathrsfs}
\usepackage{bbold}
\usepackage{dsfont}
\usepackage{natbib}
\usepackage[T1]{fontenc}
\usepackage[utf8]{inputenc}
\usepackage[colorlinks=true,linkcolor=blue,citecolor=blue, urlcolor=blue,bookmarks=false]{hyperref}

\begin{document}
\title{Energy Spectrum Theory of  Incommensurate Systems}
\author{Zhe He}
\affiliation{School of Physics and Wuhan National High Magnetic Field Center,
Huazhong University of Science and Technology, Wuhan 430074,  China}
\author{Xin-Yu Guo}
\affiliation{School of Physics and Wuhan National High Magnetic Field Center,
Huazhong University of Science and Technology, Wuhan 430074,  China}
\author{Zhen Ma}
\affiliation{School of Physics and Wuhan National High Magnetic Field Center,
Huazhong University of Science and Technology, Wuhan 430074,  China}

\author{Jin-Hua Gao}
\email{jinhua@hust.edu.cn}
\affiliation{School of Physics and Wuhan National High Magnetic Field Center,
Huazhong University of Science and Technology, Wuhan 430074,  China}


\begin{abstract}
Due to the lack of the translational symmetry,  calculating the energy spectrum of an incommensurate system has always been a theoretical challenge. Here, we propose a natural approach to generalize the energy band theory  to the incommensurate systems without reliance on the commensurate approximation, thus providing a comprehensive energy spectrum theory of the incommensurate systems. Except for a truncation dependent weighting factor, the formulae of this theory are formally  almost identical  to that of the Bloch electrons, making it particularly suitable for complex incommensurate structures.  To illustrate the application of this theory, we give three typical examples: one-dimensional bichromatic and trichromatic incommensurate potential model, as well as a moir\'{e} quasicrystal.  Our theory  establishes a fundamental  framework for understanding the incommensurate systems.
\end{abstract}
\maketitle

\emph{Introduction.}---Incommensurate (or quasiperiodic) systems refer to the quantum waves existing in multiple periodic potentials with incommensurate periods. The celebrated examples include the  Aubry-Andr\'{e}-Harper (AAH) model~\cite{AAHmodel,Harpermodel}, moir\'{e} heterostructures~\cite{Macdonald2011,cao2018correlated,cao2018SC,Lu2019,chenshaowen2020,Xu20213,young2020,2019arXiv190308130L,cao2019,Shen2020,Cai2023,Tang2020,Regan2020,zhangchendong2017,panyi2017,trilayergraphene2023}, such as twisted bilayer graphene~\cite{Macdonald2011,cao2018correlated,cao2018SC,Lu2019}, as well as certain types of  quasicrystal~\cite{quasicrystalbook,quasicrystalreview2004,plasmon2013, graphenequasicrystal2018,matter2019,Ye2020,zhoushuyun2018}. Distinct from the Bloch electrons, the incommensurate periodic potentials can lead to many unique phenomena, e.g.~wave function localization~\cite{niu2001,xie1988,boers2007,lixiao2017,bloch2018,yaohepeng2019,wangyunfei2022,xiaoteng2021,kohlert2019,lee2019,guohao2017}, moir\'{e} flat bands and superconductivity~\cite{Zhang20232,Macdonald2011,Wudoublebilayer2,Ma2021,rademaker2020prr,Koshino20191,jung20191,PhysRevX.9.031021,doi:10.1021/acs.nanolett.9b05117,jj2020prb,mazhen2023,wu20182,ashvin2019,guohuaiming2018,kennes2018,meanfield2018,xiefang2020,torma2020,kohn2019,wuxianxin2020,lianbiao2019,wangqianghua2019,youyizhuang2019,roy2019,kangjian2019,zhangyahui2019,liuchengcheng2018,khalaf2021,zhangyi2020,yutao2021}, quasicrystal electronic states with special rotational symmetries forbidden in periodic lattices~\cite{moonquasicrystal2019,helin2019,yuan2019,suzuki2019,spur2019,yangfan2023}, thereby attracting great research interest. 

With the rapid technological advancement in recent years, an increasing number of exotic incommensurate systems have been proposed and fabricated in experiments~\cite{trilayergraphene2023,Park20213,Carr20203,mora20193,zhuziyan2020,zuo20183,wangke2021,liyuhao2022,liangmiao2022,peeters2020,wanglujun2019,wangyi2019,khalaf2019,double2023,ding2023,Park20224,koshino2021,Maodan2021,yuanprb2023,guineaprb2020,shijingtian2021}, which pose significant challenges to theory. The primary difficulty  is that, due to  the lack of overall translation symmetry, the Bloch theorem becomes invalid for the incommensurate systems. Especially for complex incommensurate structures, e.g.~twisted trilayer graphene with two arbitrary twist angles~\cite{trilayergraphene2023,zhuziyan2020}, even the commonly used commensurate approximation is no longer applicable. Therefore, a comprehensive energy spectrum theory for incommensurate systems, which avoids reliance on the commensurate approximation, is urgently required~\cite{zhangpingwen2014,luskin2017,massatt2018,planewave2019,zhuziyan2020}.



In this work, we propose a natural approach to  generalize the energy band theory of the Bloch electrons to the incommensurate systems, by which the energy spectrum of the incommensurate systems can be conveniently calculated without the need for any commensurate approximation.  The formulae of such incommensurate energy spectrum (IES) theory are formally  almost identical to those of Bloch electrons, with the only difference being a truncation-dependent weighting factor for each eigenstate. Therefore, it provides a unified method for handling both commensurate and incommensurate potentials, making it an ideal choice for multiple periodic potential models.
The IES theory  establishes a fundamental theoretical framework for comprehending incommensurate systems.

\emph{Incommensurate energy spectrum theory.}---The one dimensional (1D) bichromatic incommensurate potential (BIP) model is the simplest incommensurate model~\cite{niu2001,boers2007,lixiao2017,bloch2018,yaohepeng2019}, which thus should be the best toy model to illustrate the idea of the IES theory.

The Hamiltonian of the BIP model is 
\begin{equation}\label{bichomaticmodel}
    H = -\frac{\hbar^2}{2m} \nabla^2 + \frac{V_1}{2} \cos (G_1 x) + \frac{V_2}{2} \cos (G_2 x + \phi),
\end{equation}
where $V_{1,2}$  are the amplitude of the two periodic potentials.  $G_{1,2}$ are the magnitude of the  reciprocal lattice vectors.  
We define  $\alpha=G_2 / G_1$ as the ratio between the two periods, which  is an irrational number  in the incommensurate case. 

Using plane wave basis, we get the Schr\"{o}dinger equation in momentum space
\begin{widetext}
\begin{equation}\label{centralequation}
    \dfrac{\hbar^2 q^2}{2m}  c_q + \frac{V_1}{4} c_{q-G_1} + \frac{V_1}{4} c_{q+G_1} + \frac{V_2}{4} e^{i\phi} c_{q-G_2} + \frac{V_2}{4} e^{-i\phi} c_{q+G_2} = \varepsilon c_q.
\end{equation}
\end{widetext}
Here, $c_q$ is the coefficient of the plane wave with  $\varphi (x) = \sum_q c_q e^{iqx}$, and $\varepsilon$ is the eigenvalue to be determined.  
The Eq.~\eqref{centralequation} represents a set of algebra equations,  which couple only the  plane wave states in the set   
\begin{equation}\label{qk}
  Q_{q}=\left\{ k | k=q+m G_1+nG_2 : m,n\in \mathbb{Z} \right\}. 
\end{equation}
To obtain the energy spectrum, some kind of truncation of $(m, n)$  should be given first, so that the Hamiltonian $H(q)$ becomes a  matrix of finite dimensions.  Then, for a given $q$, we can directly calculate the eigenstates of the Hamiltonian matrix $H(q)$.   Now, the key question  becomes how to determine  range of values for $q$, so that  the entire energy spectra of the incommensurate system can be obtained without any omissions or duplications.

To solve this question, let us first review the case of Bloch electrons, i.e.~$V_2=0$. In this case, we can also get a set of coupled equations as Eq.~\eqref{centralequation}, which are exactly the so-called  central equations of the Bloch electrons~\cite{solidstatebook}.   Here, the coupled plane waves are $ Q^B_{q}=\left\{ k | k=q+m G_1: m\in \mathbb{Z} \right\}$ now. In principle, $q \in (-\infty, +\infty)$. But, to calculate the eigenstates of $H(q)$, some momentum $q$ are equivalent or duplicated. It is because that the two momenta $q$ and $q+mG_1$ ($m \in \mathbb{Z}$) actually correspond to the same set of central equations. In other words, $H(q)$ and $H(q+mG_1)$ are the same matrix, thereby yielding identical energy spectra. In this sense, we say that all the momenta in $Q^B_q$  are equivalent. From such equivalence relations, we can get two important facts: 
\begin{enumerate}
\item For any arbitrary momentum q, there exists an integer $m_0$ that yields an equivalent momentum $q_0 = q + m_0G_1$ within the range $[0, G_1)$.
This  implies that the complete energy spectra can be obtained by considering only momenta within $[0, G_1)$.
\item Any two momenta in the interval $[0, G_1)$ are unequivalent. The implication is that the energy spectra are not computed repetitively while q traverses the interval $[0, G_1)$.
\end{enumerate}
Thus, the conclusion is that,  by traversing the interval $[0,G_1)$ with $q$, we can calculate the eigenstates of all $H(q)$ and obtain the complete energy spectra of the periodic system without repetition.  Obviously, this is exactly the standard procedure of the  energy band theory of the Bloch electrons, and the interval $[0, G_1)$ is  just the first Brillouin Zone (FBZ). 

\begin{figure}[tbp!]
    \centering
    \includegraphics[width=0.5\textwidth]{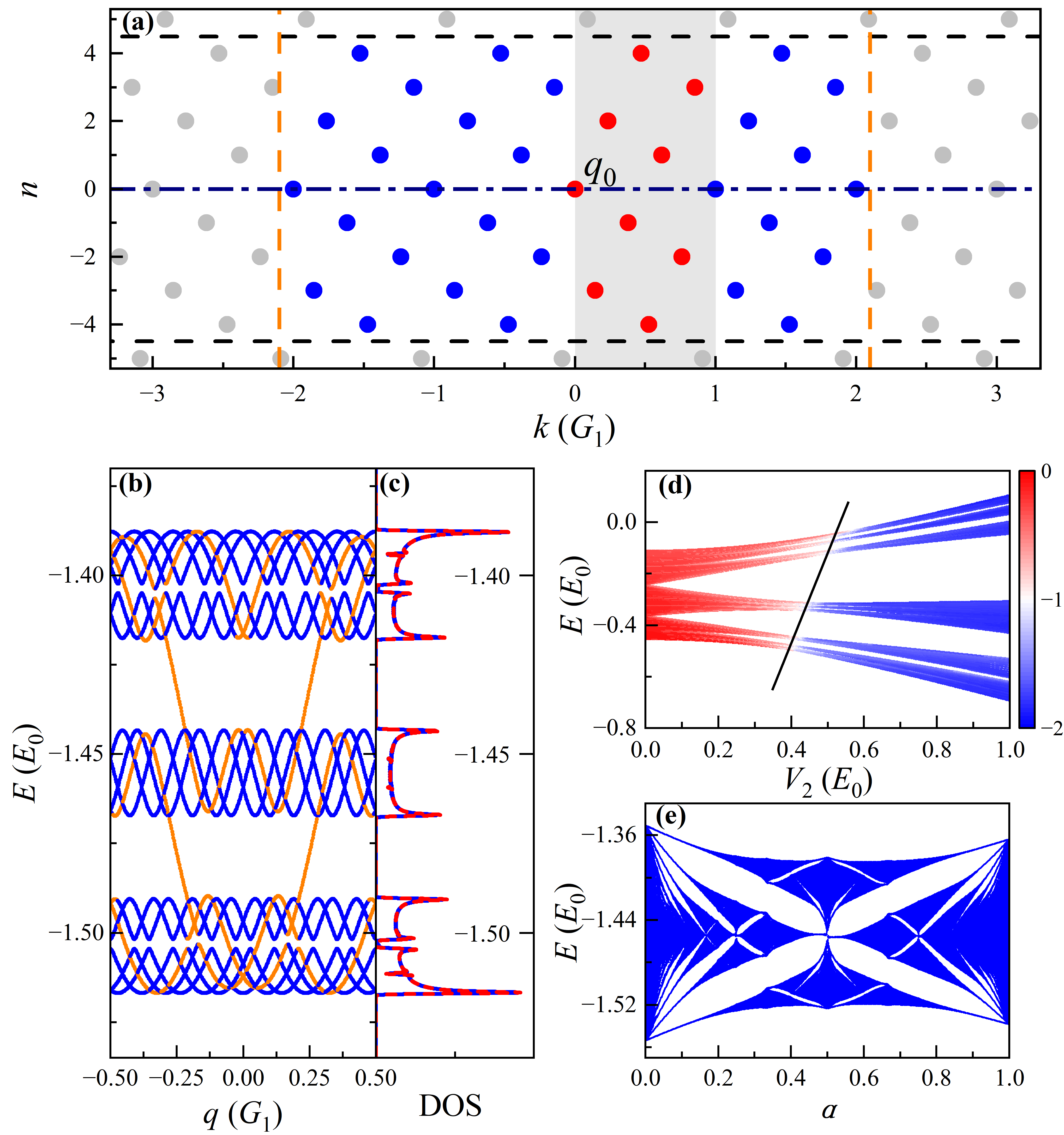}
    \caption{The BIP model with $\alpha=(\sqrt{5}-1)/2$.  (a) is the schematic of the equivalent momenta for a given $q$. We set $q=0$ here. (b) and (c) are the energy spectrum diagram and DOS. Orange lines in (b) denote the momentum edge states. Blue lines in (c) represent the DOS got from the IES method, while red lines are from the commensurate approximation $\alpha \approx 55/89$.  Parameters: $V_1=8E_0$, $V_2=0.06E_0$, $n_c=8$. (d) is the IPRM for all the eigenstates. Color represent $\log_{10} \mathrm{IPRM}$. Parameters: $V_1=4E_0$,  $n_c=40$. (e) is the butterfly like spectrum. Parameters: $V_1=8E_0$, $V_2=0.08E_0$,  $n_c=12$.  The other parameters are $\phi=0$, $k_c=4G_1$ and  $E_0 = \frac{\hbar^2}{2m} \left(\frac{G_1}{2}\right)^2$ is the energy unit. }
    \label{fig1}
\end{figure}

The similar idea can be generalized to the incommensurate case. As indicated in Eq.~\eqref{centralequation} and \eqref{qk}, all the momenta in $Q_{q}=\left\{ k | k=q+m G_1+nG_2 : m,n\in \mathbb{Z} \right\}$ are equivalent.
Such equivalence relation can also give us several important messages:
\begin{enumerate}
\item With a specified $n$, we can always find an equivalent momentum $q_n=(q+nG_2)+m_n G_1$ within the range of $[0,G_1)$ for any given $q$, where $m_n$ is an integer.  
\item Suppose a truncation of $n$ is provided, and $N_E$ represents the total number of all the allowed $n$. Then, for any $q$, there are $N_E$ equivalent momenta in  the interval $[0, G_1)$, which do not overlap with each other due to the incommensurability. 
\end{enumerate}
Fig.~\ref{fig1} (a) can help us intuitively understand the two points above. The vertical (horizontal) axis of Fig.~\ref{fig1} (a) denotes value of $n$ ($k$), 
so that all equivalent momenta $k = q + mG_1 + nG_2$ for a given $q$ can be represented as discrete dots on the graph. Moreover, for a specified $n$, the equivalent momenta $(q+n G_2)+mG_1$ with different values of $m$ lie on a horizontal line. Within each horizontal line, we see that there always exists one equivalent momentum  within the range $[0, G_1)$ (red dot), which is just the $q_n$ mentioned above.  A natural truncation scheme of $(m,n)$ is: (1) $|n| \leq n_c$; (2) $|k| \leq k_c$. Here, $n_c$ and $k_c$ are two truncation constants, where $k_c$ reflects the truncated energy of the plane waves and $n_c$ determines the minimum interval between all the coupled  momenta~\footnote{See Supplemental Material at [URL]  for the interpretation about the meaning of $n_c$.}.  The truncation of $n$ ($k$) is represented as the  two black horizontal  (orange vertical) dashed lines in Fig.~\ref{fig1} (a). Once $n_c$ is given, the total number of allowed $n$ (red dots) are $N_E=2 n_c +1$. 
It can be proved that all the red dots ($q_n$) never overlap with each other in the incommensurate case~\footnote{See Supplemental Material at [URL] for the proof of this statment.}.
Therefore, for any $q$, we can always find $N_E$ distinct equivalent momenta in the interval $[0, G_1)$. 



The two points above suggest that  a correct and convenient way to calculate the  energy spectra of an incommensurate system is to let $q$ traverse the interval $[0,G_1)$, calculate the eigenstates of each $H(q)$ and assign a weight factor of $1/N_E$ to each eigenstate. 

Based on this idea, the density of states (DOS) and expectation value of an observable in an incommensurate system can be calculated by the following formulae:
\begin{align}
   \label{DOS}  \rho (\varepsilon) = & \frac{1}{N_E} \sum_{q \in [0,G_1),i} \delta (\varepsilon - \varepsilon_{qi})\\
  \label{expectation} \langle \hat{A} \rangle =  &  \frac{1}{N_E}  \sum_{q \in [0,G_1),i} p_{qi} \langle \phi_{qi} |\hat{A}| \phi_{qi} \rangle
\end{align}
where $\varepsilon_{qi}$ and $ | \phi_{qi} \rangle$ denote  the $i$th eigenvalue and eigenstate of $H(q)$, and $p_{qi}$ is the Boltzmann factor. Here, we name the interval $[0, G_1)$ primary Brillouin zone (PBZ) of the incommensurate systems. Note that we can also choose $G_2$ as the PBZ, which will give the same results with proper truncation~\footnote{See Supplemental Material at [URL]  for the calculation results with $G_2$ as PBZ.}.
 The two formulae above are formally almost identical to that of Bloch electrons, which are the central results of the proposed IES theory.

The IES method can provide accurate enough numerical results as long as the  cutoff $k_c$ and $n_c$ is large enough,  since the plane wave basis is used.
In practical calculations, we need to test different truncation values to ensure the convergence of results.  An interesting case is $n_c \to \infty$, where the equivalent momenta $q_n$ of an incommensurate model will uniformly and densely fill the whole PBZ. It implies that,  once $n_c$ is sufficiently large, the entire incommensurate spectra can be obtained by just calculating the eigenstates of $H(q)$ with one any chosen $q$ ~\cite{zhuziyan2020,planewave2019}. But the cost is  the dimension of the Hamiltonian matrix is extremely huge.

The IES method, i.e.~Eq.\eqref{DOS} and \eqref{expectation}, is also valid  for the commensurate case. The  only difference is that the $N_E$ for the commensurate case has to be determined by directly counting the distinct $q_n$ in PBZ,   since  the the equivalent momenta in the PBZ ($q_n$)   now  may overlap with one another~\footnote{See Supplemental Material at [URL] for the example of commensurate case.}. In this sense, the incommensurate and commensurate systems exhibit  no fundamental distinction, which thus can be  treated  under a unified theoretical framework.


\emph{Results of the BIP model.}---We then apply this IES method to three specific examples for demonstration. The first one is the BIP model, where we set $\alpha=(\sqrt{5}-1)/2$. It is a typical incommensurate case and has been intensively studied with various methods~\cite{niu2001,boers2007,lixiao2017,bloch2018,yaohepeng2019}.

\begin{figure}[tbp!]
    \centering
    \includegraphics[width=0.5\textwidth]{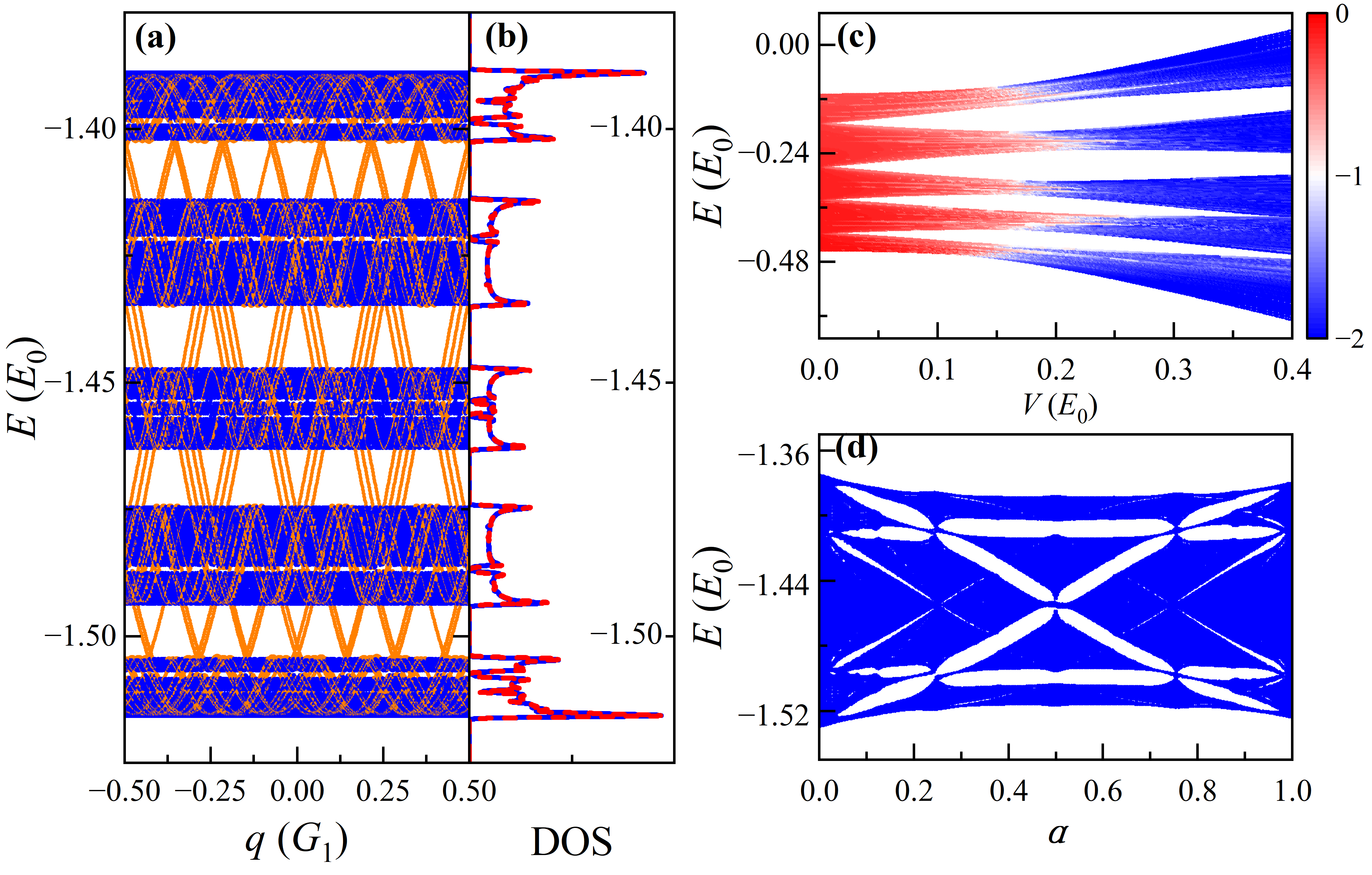}
    \caption{The TIP model.   (a) and (b) show the energy spectrum diagram and DOS. Orange lines in (a) denote the momentum edge states. Blue lines in (b) represent the DOS got from the IES method, while red lines are from the commensurate approximation $G_1:G_2:G_3 \approx 65:49:37$. Parameters: $V_1=8E_0$, $V_2=V_3=0.03E_0$, $n_c=6$. (c) is the IPRM for all the eigenstates. Colors represent $\log_{10} \mathrm{IPRM}$ and we set $V_2=V_3=V$. Parameters: $V_1=4E_0$,  $n_c=8$. (d) is the butterfly like spectrum. Parameters: $V_1=8E_0$, $V_2=V_3=0.03E_0$,  $n_c=8$, $\alpha_2=\lambda^{-1}$, $\alpha_3=\alpha$.  The other parameters are $\phi=0$, $k_c=4G_1$ and  $E_0 = \frac{\hbar^2}{2m} \left(\frac{G_1}{2}\right)^2$ is the energy unit. The two irrational numbers are $\alpha_2=\lambda^{-1}$, $\alpha_3=\lambda^{-2}$, where $\lambda=1.3247 \cdots$ is the root of the equation $x^3-x-1=0$~\cite{TIP1989}.}
    \label{fig2}
\end{figure}

The numerical results are given in Fig.~\ref{fig1}.   
First of all, like the Bloch electrons, we can plot energy spectra $\varepsilon_{qi}$ as a function of $q$ [see Fig.~\ref{fig1} (b)], where $i$ become the ``band'' index and $q \in \mathrm{PBZ}$. Such energy spectrum diagram clearly indicates the existence of the energy gaps.  However, the incommensurate energy spectrum diagram here is different from the energy bands of Bloch electrons, because its shape depends on the $N_E$, i.e. the truncation $n_c$. Note that, in the PBZ, there are $N_E$ identical spectra at the $N_E$ equivalent momenta, which is quite unlike that of Bloch electrons.   The other obvious characteristic is that there exist a kind of momentum edge states (orange lines) in the  energy gaps.  Formally, Eq.~\eqref{centralequation} can be viewed as  a tight binding (TB) model in momentum space, where a plane wave is equivalent to  a discrete site in momentum space. Thus, when the truncation $n_c$ is given, it actually results in open boundaries in momentum space. Like the TB model in real space, the open boundary will give rise to edge states, i.e. the so-called momentum edge states in Fig.~\ref{fig1} (b). We can distinguish the momentum edge states by examining the distribution of the wave functions in momentum space~\footnote{See Supplemental Material at [URL] for the details about momentum edge states.}.  Since the momentum edge states rely on the truncation $n_c$, we think that they are artificially induced states that need to be eliminated.

To demonstrate the correctness of the IES method, we compare it with the commensurate approximation. The DOS calculated with the two methods are plotted in Fig.~\ref{fig1} (c), where the blue (red) lines represent the IES method (commensurate approximation with $\alpha \approx 55/89$). Both methods yield the same DOS, which clearly indicates that the IES method can correctly  describe the energy spectrum of the incommensurate system.

 The IES method can accurately capture the wave function characteristics of the incommensurate system as well. A direct example is the single particle mobility edge of the BIP model,  which refers to the quantum localization of the wave functions in an incommensurate system. With the IES method, the inverse participation ratio in momentum space (IPRM) can be utilized to quantify the degree of localization~\cite{localization2021},
\begin{equation}
    \mathrm{IPRM} (|\phi_{qi}\rangle) =\sum_{m,n} |c_{q+mG_1+nG_2}|^4.
\end{equation}
If the wave function is localized in real space, it should be extended in momentum space, and thus the IPRM will approach to zero, i.e.~$\log_{10}(\mathrm{IPRM}) \to - \infty$,  as $k_c$ increases. Conversely, for an extended wave function, $\log_{10}(\mathrm{IPRM})$ tends to zero. Fig.~\ref{fig1} (e) plots the IPRM of all the eigenstates of the BIP model as a function of $V_2$, where the  $\log_{10} (\mathrm{IPRM})$ is represented by the color. Clearly, when $V_2$ is small (large), the wave function is extended (localized). Moreover, we can see that the localization transition is energy dependent (the black oblique line), which indicates that the single particle mobility edge indeed exists. The results from the IES method are completely consistent with the known conclusions~\cite{lixiao2017,bloch2018}.

Very interestingly,  the IES method offers an improved approach to calculate the famous Hofstadter butterfly spectrum.   It is known that in the TB limit (with deep $V_1$ potential), the BIP model will asymptotically approach the AAH model, which can be exactly mapped into a 2D Hofstadter model, i.e.~2D square lattice in a perpendicular magnetic field~\cite{caixiaoming2011}. Thus,  when the energy spectrum of the BIP model is plotted as a function of $\alpha$, we indeed get a butterfly-like spectrum, as shown in Fig.~\ref{fig1} (e).  Since  $\alpha$  is proportional to the magnetic flux $\phi$ through each plaquette, it means that the IES theory provides a way to  calculate the Hofstadter butterfly spectrum under any magnetic field, no matter whether $\phi$ is rational or irrational.

\begin{figure}[tbp!]
    \centering
    \includegraphics[width=0.5\textwidth]{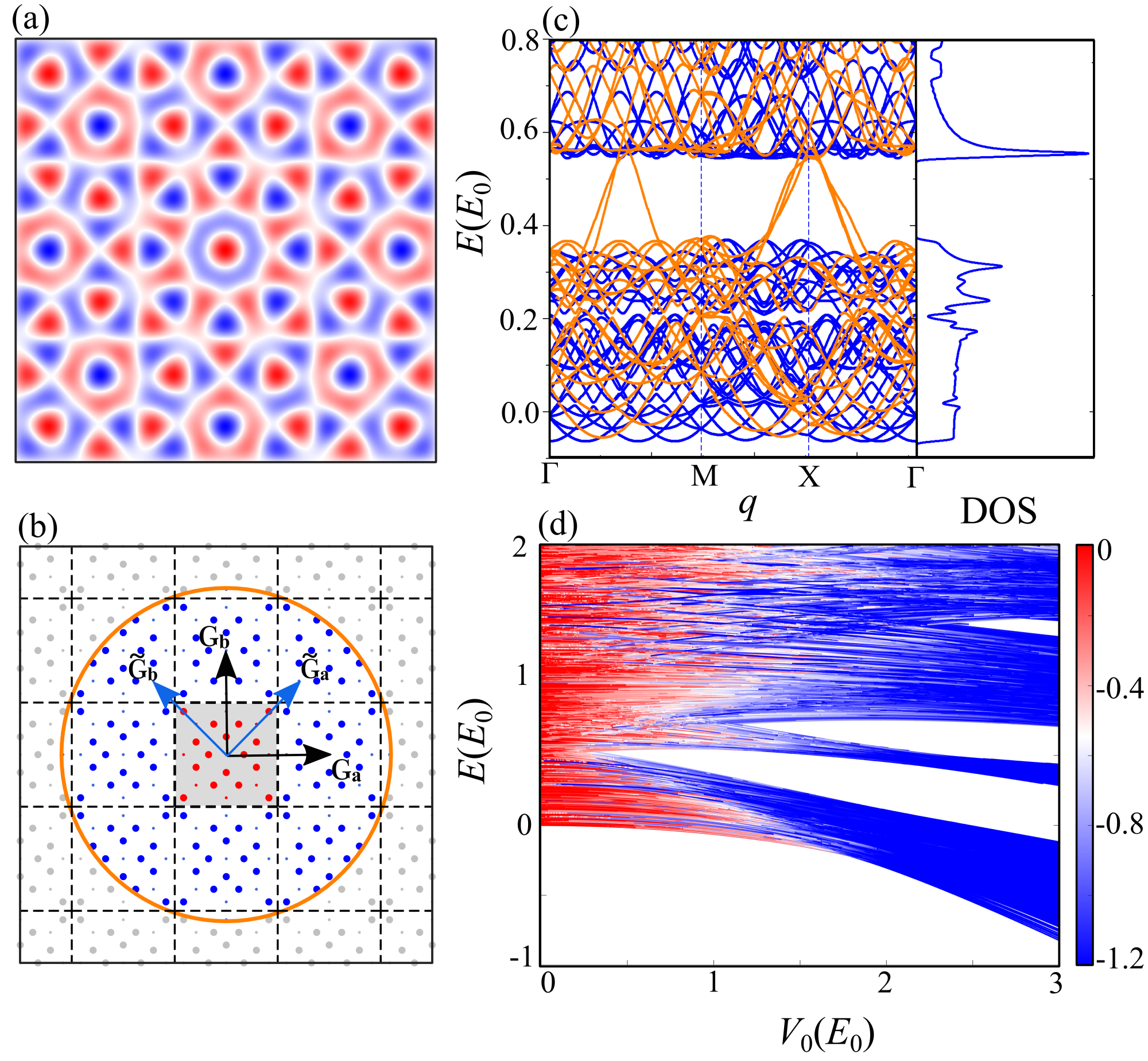}
    \caption{Moir\'{e} quasicrystal. (a) shows the potential of moir\'{e} quasicrystal. (b) is the schematic of the equivalent momenta. (c) is the energy spectrum and the corresponding DOS. Parameters: $V_0=1E_0$, $n_c=4$ and $k_c=2.1 |\mathbf{G_a}|$. (d) shows the IPRM, where the color represents $\log_{10} (\mathrm{IPRM})$. 
    Parameters:  $n_c=7$ and $k_c=2.1 |\mathbf{G_a}|$. $E_0 = 2\frac{\hbar^2}{2m} \left(\frac{|\mathbf{G_a}|}{2}\right)^2$ is the energy unit.}
    \label{fig3}
\end{figure}

\emph{Results of the TIP model.}---The IES method offers a convenient approach for calculating complex incommensurate systems.  One example is the 1D trichromatic incommensurate potential (TIP) model, where three applied incommensurate potentials greatly hinder the commensurate approximation.  The Hamiltonian is now
\begin{equation}
    H = -\frac{\hbar^2}{2m} \nabla^2 + \sum_{i=1,2,3} \frac{V_i}{2} \cos (G_i x).
\end{equation}
where $\alpha_2=G_2/G_1$ and $\alpha_3=G_3 / G_1$ are two irrational numbers~\cite{TIP1989}.  The coupled momenta are  now  $Q_{q}=\left\{ k | k=q+m G_1+n_2G_2 + n_3 G_3: m,n_2,n_3\in \mathbb{Z} \right\}$. With the IES theory,  we can still choose $[0,G_1)$ as the PBZ and use the truncation:  $|n_{2,3}| \leq n_{c}$ and $|k| \leq k_c$. Similarly, we also have $N_E$ equivalent momenta in the PBZ, which can be determined by directly enumerating the distinct equivalent momenta in PBZ. Then, energy spectrum and other properties can be calculated with Eq.~\eqref{DOS} and \eqref{expectation} in the same way.  

The numerical results are given in Fig.~\ref{fig2}. Fig.~\ref{fig2} (a) shows the energy spectrum. With three incommensurate potentials, the number of equivalent momenta $N_E$ is markedly greater than that in the BIP model, resulting in a highly dense energy spectrum. 
The presence of momentum edge states is also observed in this case (orange lines). We have calculated the corresponding DOS (blue lines) in Fig.~\ref{fig2} (b), which coincides well with that of an approximate commensurate structure (red lines). In Fig.~\ref{fig2} (d), we present the IPRM to show the localization feature of the eigenstates. We also can get a butterfly-like spectrum by plotting the energy spectra as a function of $\alpha_3$ with a fixed $\alpha_2$, see Fig.~\ref{fig2} (e). 

\emph{Results of moir\'{e} quasicrystal.}---The two dimensional (2D) incommensurate systems have drawn great interest very recently. Here, we use the IES method to calculate a special 2D incommensurate system, i.e.~moir\'{e} quasicrystal, which has eightfold rotational symmetry but no translation symmetry.   Very similar to the BIP model, such moir\'{e} quasicrystal can be achieved by applying two square periodic potentials $V(\theta_1=0)$ and $V(\theta_2=\pi/4)$ with a twist angle $\theta=\pi/4$, where 
\begin{equation}
  V(\theta)=\frac{V_0}{4} \{ \cos[R(\theta)\mathbf{G_a}\cdot\bm{r}] + \cos[R(\theta)\mathbf{G_b}\cdot\bm{r}]\}.   
\end{equation}
$\mathbf{G_a}$ and $\mathbf{G_b}$ are the two reciprocal lattice vectors of a square lattice, and $R(\theta)$ is the rotation matrix. The  moir\'{e} quasicrystal  potential is plotted in Fig.~\ref{fig3} (a),  revealing its evident eightfold rotational symmetry. Note that such moir\'{e} quasicrystal has already been realized in cold atom systems~\cite{matter2019}\footnote{See Supplemental Material at [URL] for the relation to the experimental model.}.

For a given $\bm{q}$, the  coupled momenta  are now $\mathbf{k}=\mathbf{q}+m_1 \mathbf{G_a}+m_2 \mathbf{G_b} + n_1 \mathbf{\Tilde{G}_a} + n_2 \mathbf{\Tilde{G}_b}$, where we define $\mathbf{\Tilde{G}_{a,b}} \equiv R(\pi/4) \mathbf{G_{a,b}}$ and $m_{1,2}, n_{1,2} \in \mathbb{Z}$. The equivalent momenta and the truncation are illustrated in Fig.~\ref{fig3} (b). Here, we  choose the FBZ of  $V(\theta_1=0)$ as the PBZ, see the gray square. All the equivalent momenta are plotted as dots in Fig.~\ref{fig3} (b), with a truncation:   $|n_{1,2}| \leq n_c$, $|k| \leq k_c$. $k_c$ denotes the  energy cutoff of the plane waves,  illustrated as the orange circle.  For the moir\'{e} quasicrystal, all the equivalent momenta in PBZ will never overlap with one another (red dots)~\footnote{See Supplemental Material at [URL] for the proof of this statement.}, which implies $N_{E}=(2n_c+1)^2$.  The calculated DOS is given in Fig.~\ref{fig3} (c).   In Fig.~\ref{fig3} (d), we plot the IPRM of the moir\'{e} quasicrystal.  The IES theory does provide a convenient method for handling the moir\'{e} quasicrystal. All the calculation details of the three examples are given in the supplementary materials~\footnote{See Supplemental Material at [URL] for the calculation details.}.

\emph{Summary.}---In summary, we establish a general energy spectrum theory for the incommensurate systems without relying on the commensurate approximation. In addition to the examples above, it can be readily applied to the TB models and other artificial incommensurate systems, e.g.~cold atom systems and optical crystals. More importantly, it implies that all the formulae of the energy band theory can be directly transplanted into the incommensurate systems via the IES theory. Thus, our theory actually gives a fundamental framework for comprehending the incommensurate systems.



\begin{acknowledgments}
This work was supported by the National Key Research and Development Program of China (No.~2022YFA1403501), and the National Natural Science Foundation of China(Grants No.~12141401, No.~11874160).
\end{acknowledgments}

	\bibliography{references}

\begin{thebibliography}{107}%
\makeatletter
\providecommand \@ifxundefined [1]{%
 \@ifx{#1\undefined}
}%
\providecommand \@ifnum [1]{%
 \ifnum #1\expandafter \@firstoftwo
 \else \expandafter \@secondoftwo
 \fi
}%
\providecommand \@ifx [1]{%
 \ifx #1\expandafter \@firstoftwo
 \else \expandafter \@secondoftwo
 \fi
}%
\providecommand \natexlab [1]{#1}%
\providecommand \enquote  [1]{``#1''}%
\providecommand \bibnamefont  [1]{#1}%
\providecommand \bibfnamefont [1]{#1}%
\providecommand \citenamefont [1]{#1}%
\providecommand \href@noop [0]{\@secondoftwo}%
\providecommand \href [0]{\begingroup \@sanitize@url \@href}%
\providecommand \@href[1]{\@@startlink{#1}\@@href}%
\providecommand \@@href[1]{\endgroup#1\@@endlink}%
\providecommand \@sanitize@url [0]{\catcode `\\12\catcode `\$12\catcode
  `\&12\catcode `\#12\catcode `\^12\catcode `\_12\catcode `\%12\relax}%
\providecommand \@@startlink[1]{}%
\providecommand \@@endlink[0]{}%
\providecommand \url  [0]{\begingroup\@sanitize@url \@url }%
\providecommand \@url [1]{\endgroup\@href {#1}{\urlprefix }}%
\providecommand \urlprefix  [0]{URL }%
\providecommand \Eprint [0]{\href }%
\providecommand \doibase [0]{http://dx.doi.org/}%
\providecommand \selectlanguage [0]{\@gobble}%
\providecommand \bibinfo  [0]{\@secondoftwo}%
\providecommand \bibfield  [0]{\@secondoftwo}%
\providecommand \translation [1]{[#1]}%
\providecommand \BibitemOpen [0]{}%
\providecommand \bibitemStop [0]{}%
\providecommand \bibitemNoStop [0]{.\EOS\space}%
\providecommand \EOS [0]{\spacefactor3000\relax}%
\providecommand \BibitemShut  [1]{\csname bibitem#1\endcsname}%
\let\auto@bib@innerbib\@empty
\bibitem [{\citenamefont {Aubry}\ and\ \citenamefont
  {André}(1980)}]{AAHmodel}%
  \BibitemOpen
  \bibfield  {author} {\bibinfo {author} {\bibfnamefont {S.}~\bibnamefont
  {Aubry}}\ and\ \bibinfo {author} {\bibfnamefont {G.}~\bibnamefont {André}},\
  }\href@noop {} {\bibfield  {journal} {\bibinfo  {journal} {Ann. Israel Phys.
  Soc.}\ }\textbf {\bibinfo {volume} {3}},\ \bibinfo {pages} {18} (\bibinfo
  {year} {1980})}\BibitemShut {NoStop}%
\bibitem [{\citenamefont {Harper}(1955)}]{Harpermodel}%
  \BibitemOpen
  \bibfield  {author} {\bibinfo {author} {\bibfnamefont {P.~G.}\ \bibnamefont
  {Harper}},\ }\href@noop {} {\bibfield  {journal} {\bibinfo  {journal} {Proc.
  Phys. Soc. A}\ }\textbf {\bibinfo {volume} {68}},\ \bibinfo {pages} {874}
  (\bibinfo {year} {1955})}\BibitemShut {NoStop}%
\bibitem [{\citenamefont {Bistritzer}\ and\ \citenamefont
  {MacDonald}(2011)}]{Macdonald2011}%
  \BibitemOpen
  \bibfield  {author} {\bibinfo {author} {\bibfnamefont {R.}~\bibnamefont
  {Bistritzer}}\ and\ \bibinfo {author} {\bibfnamefont {A.~H.}\ \bibnamefont
  {MacDonald}},\ }\href {\doibase 10.1073/pnas.1108174108} {\bibfield
  {journal} {\bibinfo  {journal} {Proc. Natl. Acad. Sci.}\ }\textbf {\bibinfo
  {volume} {108}},\ \bibinfo {pages} {12233} (\bibinfo {year}
  {2011})}\BibitemShut {NoStop}%
\bibitem [{\citenamefont {Cao}\ \emph {et~al.}(2018{\natexlab{a}})\citenamefont
  {Cao}, \citenamefont {Fatemi}, \citenamefont {Demir} \emph
  {et~al.}}]{cao2018correlated}%
  \BibitemOpen
  \bibfield  {author} {\bibinfo {author} {\bibfnamefont {Y.}~\bibnamefont
  {Cao}}, \bibinfo {author} {\bibfnamefont {V.}~\bibnamefont {Fatemi}},
  \bibinfo {author} {\bibfnamefont {A.}~\bibnamefont {Demir}},  \emph
  {et~al.},\ }\href {\doibase 10.1038/nature26154} {\bibfield  {journal}
  {\bibinfo  {journal} {Nature}\ }\textbf {\bibinfo {volume} {556}},\ \bibinfo
  {pages} {80} (\bibinfo {year} {2018}{\natexlab{a}})}\BibitemShut {NoStop}%
\bibitem [{\citenamefont {Cao}\ \emph {et~al.}(2018{\natexlab{b}})\citenamefont
  {Cao}, \citenamefont {Fatemi}, \citenamefont {Fang} \emph
  {et~al.}}]{cao2018SC}%
  \BibitemOpen
  \bibfield  {author} {\bibinfo {author} {\bibfnamefont {Y.}~\bibnamefont
  {Cao}}, \bibinfo {author} {\bibfnamefont {V.}~\bibnamefont {Fatemi}},
  \bibinfo {author} {\bibfnamefont {S.}~\bibnamefont {Fang}},  \emph {et~al.},\
  }\href {\doibase 10.1038/nature26160} {\bibfield  {journal} {\bibinfo
  {journal} {Nature}\ }\textbf {\bibinfo {volume} {556}},\ \bibinfo {pages}
  {43} (\bibinfo {year} {2018}{\natexlab{b}})}\BibitemShut {NoStop}%
\bibitem [{\citenamefont {Lu}\ \emph {et~al.}(2019)\citenamefont {Lu},
  \citenamefont {Stepanov}, \citenamefont {Yang} \emph {et~al.}}]{Lu2019}%
  \BibitemOpen
  \bibfield  {author} {\bibinfo {author} {\bibfnamefont {X.}~\bibnamefont
  {Lu}}, \bibinfo {author} {\bibfnamefont {P.}~\bibnamefont {Stepanov}},
  \bibinfo {author} {\bibfnamefont {W.}~\bibnamefont {Yang}},  \emph {et~al.},\
  }\href {\doibase 10.1038/s41586-019-1695-0} {\bibfield  {journal} {\bibinfo
  {journal} {Nature}\ }\textbf {\bibinfo {volume} {574}},\ \bibinfo {pages}
  {653} (\bibinfo {year} {2019})}\BibitemShut {NoStop}%
\bibitem [{\citenamefont {Chen}\ \emph
  {et~al.}(2021{\natexlab{a}})\citenamefont {Chen}, \citenamefont {He},
  \citenamefont {Zhang} \emph {et~al.}}]{chenshaowen2020}%
  \BibitemOpen
  \bibfield  {author} {\bibinfo {author} {\bibfnamefont {S.}~\bibnamefont
  {Chen}}, \bibinfo {author} {\bibfnamefont {M.}~\bibnamefont {He}}, \bibinfo
  {author} {\bibfnamefont {Y.-H.}\ \bibnamefont {Zhang}},  \emph {et~al.},\
  }\href {https://doi.org/10.1038/s41567-020-01062-6} {\bibfield  {journal}
  {\bibinfo  {journal} {Nat. Phys.}\ }\textbf {\bibinfo {volume} {17}},\
  \bibinfo {pages} {374} (\bibinfo {year} {2021}{\natexlab{a}})}\BibitemShut
  {NoStop}%
\bibitem [{\citenamefont {Xu}\ \emph {et~al.}(2021)\citenamefont {Xu},
  \citenamefont {Al~Ezzi}, \citenamefont {Balakrishnan} \emph
  {et~al.}}]{Xu20213}%
  \BibitemOpen
  \bibfield  {author} {\bibinfo {author} {\bibfnamefont {S.}~\bibnamefont
  {Xu}}, \bibinfo {author} {\bibfnamefont {M.~M.}\ \bibnamefont {Al~Ezzi}},
  \bibinfo {author} {\bibfnamefont {N.}~\bibnamefont {Balakrishnan}},  \emph
  {et~al.},\ }\href {\doibase 10.1038/s41567-021-01172-9} {\bibfield  {journal}
  {\bibinfo  {journal} {Nat. Phys.}\ }\textbf {\bibinfo {volume} {17}},\
  \bibinfo {pages} {619} (\bibinfo {year} {2021})}\BibitemShut {NoStop}%
\bibitem [{\citenamefont {Polshyn}\ \emph {et~al.}(2020)\citenamefont
  {Polshyn}, \citenamefont {Zhu}, \citenamefont {Kumar} \emph
  {et~al.}}]{young2020}%
  \BibitemOpen
  \bibfield  {author} {\bibinfo {author} {\bibfnamefont {H.}~\bibnamefont
  {Polshyn}}, \bibinfo {author} {\bibfnamefont {J.}~\bibnamefont {Zhu}},
  \bibinfo {author} {\bibfnamefont {M.}~\bibnamefont {Kumar}},  \emph
  {et~al.},\ }\href {\doibase https://doi.org/10.1038/s41586-020-2963-8}
  {\bibfield  {journal} {\bibinfo  {journal} {Nature}\ }\textbf {\bibinfo
  {volume} {588}},\ \bibinfo {pages} {66} (\bibinfo {year} {2020})}\BibitemShut
  {NoStop}%
\bibitem [{\citenamefont {Liu}\ \emph {et~al.}(2020)\citenamefont {Liu},
  \citenamefont {Hao}, \citenamefont {Khalaf} \emph
  {et~al.}}]{2019arXiv190308130L}%
  \BibitemOpen
  \bibfield  {author} {\bibinfo {author} {\bibfnamefont {X.}~\bibnamefont
  {Liu}}, \bibinfo {author} {\bibfnamefont {Z.}~\bibnamefont {Hao}}, \bibinfo
  {author} {\bibfnamefont {E.}~\bibnamefont {Khalaf}},  \emph {et~al.},\ }\href
  {\doibase https://doi.org/10.1038/s41586-020-2458-7} {\bibfield  {journal}
  {\bibinfo  {journal} {Nature}\ }\textbf {\bibinfo {volume} {583}},\ \bibinfo
  {pages} {221} (\bibinfo {year} {2020})}\BibitemShut {NoStop}%
\bibitem [{\citenamefont {Cao}\ \emph {et~al.}(2020)\citenamefont {Cao},
  \citenamefont {Rodan-Legrain}, \citenamefont {Rubies-Bigorda} \emph
  {et~al.}}]{cao2019}%
  \BibitemOpen
  \bibfield  {author} {\bibinfo {author} {\bibfnamefont {Y.}~\bibnamefont
  {Cao}}, \bibinfo {author} {\bibfnamefont {D.}~\bibnamefont {Rodan-Legrain}},
  \bibinfo {author} {\bibfnamefont {O.}~\bibnamefont {Rubies-Bigorda}},  \emph
  {et~al.},\ }\href {\doibase https://doi.org/10.1038/s41586-020-2260-6}
  {\bibfield  {journal} {\bibinfo  {journal} {Nature}\ }\textbf {\bibinfo
  {volume} {583}},\ \bibinfo {pages} {215} (\bibinfo {year}
  {2020})}\BibitemShut {NoStop}%
\bibitem [{\citenamefont {Shen}\ \emph {et~al.}(2020)\citenamefont {Shen},
  \citenamefont {Chu}, \citenamefont {Wu} \emph {et~al.}}]{Shen2020}%
  \BibitemOpen
  \bibfield  {author} {\bibinfo {author} {\bibfnamefont {C.}~\bibnamefont
  {Shen}}, \bibinfo {author} {\bibfnamefont {Y.}~\bibnamefont {Chu}}, \bibinfo
  {author} {\bibfnamefont {Q.}~\bibnamefont {Wu}},  \emph {et~al.},\ }\href
  {\doibase 10.1038/s41567-020-0825-9} {\bibfield  {journal} {\bibinfo
  {journal} {Nat. Phys.}\ }\textbf {\bibinfo {volume} {16}},\ \bibinfo {pages}
  {520} (\bibinfo {year} {2020})}\BibitemShut {NoStop}%
\bibitem [{\citenamefont {Cai}\ \emph {et~al.}(2023)\citenamefont {Cai},
  \citenamefont {Anderson}, \citenamefont {Wang} \emph {et~al.}}]{Cai2023}%
  \BibitemOpen
  \bibfield  {author} {\bibinfo {author} {\bibfnamefont {J.}~\bibnamefont
  {Cai}}, \bibinfo {author} {\bibfnamefont {E.}~\bibnamefont {Anderson}},
  \bibinfo {author} {\bibfnamefont {C.}~\bibnamefont {Wang}},  \emph {et~al.},\
  }\href {\doibase 10.1038/s41586-023-06289-w} {\bibfield  {journal} {\bibinfo
  {journal} {Nature}\ } (\bibinfo {year} {2023}),\
  10.1038/s41586-023-06289-w}\BibitemShut {NoStop}%
\bibitem [{\citenamefont {Tang}\ \emph {et~al.}(2020)\citenamefont {Tang},
  \citenamefont {Li}, \citenamefont {Li} \emph {et~al.}}]{Tang2020}%
  \BibitemOpen
  \bibfield  {author} {\bibinfo {author} {\bibfnamefont {Y.}~\bibnamefont
  {Tang}}, \bibinfo {author} {\bibfnamefont {L.}~\bibnamefont {Li}}, \bibinfo
  {author} {\bibfnamefont {T.}~\bibnamefont {Li}},  \emph {et~al.},\ }\href
  {\doibase 10.1038/s41586-020-2085-3} {\bibfield  {journal} {\bibinfo
  {journal} {Nature}\ }\textbf {\bibinfo {volume} {579}},\ \bibinfo {pages}
  {353} (\bibinfo {year} {2020})}\BibitemShut {NoStop}%
\bibitem [{\citenamefont {Regan}\ \emph {et~al.}(2020)\citenamefont {Regan},
  \citenamefont {Wang}, \citenamefont {Jin} \emph {et~al.}}]{Regan2020}%
  \BibitemOpen
  \bibfield  {author} {\bibinfo {author} {\bibfnamefont {E.~C.}\ \bibnamefont
  {Regan}}, \bibinfo {author} {\bibfnamefont {D.}~\bibnamefont {Wang}},
  \bibinfo {author} {\bibfnamefont {C.}~\bibnamefont {Jin}},  \emph {et~al.},\
  }\href {\doibase 10.1038/s41586-020-2092-4} {\bibfield  {journal} {\bibinfo
  {journal} {Nature}\ }\textbf {\bibinfo {volume} {579}},\ \bibinfo {pages}
  {359} (\bibinfo {year} {2020})}\BibitemShut {NoStop}%
\bibitem [{\citenamefont {Zhang}\ \emph {et~al.}(2017)\citenamefont {Zhang},
  \citenamefont {Chuu}, \citenamefont {Ren} \emph
  {et~al.}}]{zhangchendong2017}%
  \BibitemOpen
  \bibfield  {author} {\bibinfo {author} {\bibfnamefont {C.}~\bibnamefont
  {Zhang}}, \bibinfo {author} {\bibfnamefont {C.-P.}\ \bibnamefont {Chuu}},
  \bibinfo {author} {\bibfnamefont {X.}~\bibnamefont {Ren}},  \emph {et~al.},\
  }\href {\doibase 10.1126/sciadv.1601459} {\bibfield  {journal} {\bibinfo
  {journal} {Sci. Adv.}\ }\textbf {\bibinfo {volume} {3}},\ \bibinfo {pages}
  {e1601459} (\bibinfo {year} {2017})}\BibitemShut {NoStop}%
\bibitem [{\citenamefont {Tong}\ \emph {et~al.}(2017)\citenamefont {Tong},
  \citenamefont {Yu}, \citenamefont {Zhu} \emph {et~al.}}]{panyi2017}%
  \BibitemOpen
  \bibfield  {author} {\bibinfo {author} {\bibfnamefont {Q.}~\bibnamefont
  {Tong}}, \bibinfo {author} {\bibfnamefont {H.}~\bibnamefont {Yu}}, \bibinfo
  {author} {\bibfnamefont {Q.}~\bibnamefont {Zhu}},  \emph {et~al.},\ }\href
  {\doibase 10.1038/nphys3968} {\bibfield  {journal} {\bibinfo  {journal} {Nat.
  Phys.}\ }\textbf {\bibinfo {volume} {13}},\ \bibinfo {pages} {356} (\bibinfo
  {year} {2017})}\BibitemShut {NoStop}%
\bibitem [{\citenamefont {Uri}\ \emph {et~al.}(2023)\citenamefont {Uri},
  \citenamefont {de~la Barrera}, \citenamefont {Randeria} \emph
  {et~al.}}]{trilayergraphene2023}%
  \BibitemOpen
  \bibfield  {author} {\bibinfo {author} {\bibfnamefont {A.}~\bibnamefont
  {Uri}}, \bibinfo {author} {\bibfnamefont {S.}~\bibnamefont {de~la Barrera}},
  \bibinfo {author} {\bibfnamefont {M.}~\bibnamefont {Randeria}},  \emph
  {et~al.},\ }\href {\doibase 10.1038/s41586-023-06294-z} {\bibfield  {journal}
  {\bibinfo  {journal} {Nature}\ }\textbf {\bibinfo {volume} {620}},\ \bibinfo
  {pages} {1} (\bibinfo {year} {2023})}\BibitemShut {NoStop}%
\bibitem [{\citenamefont {Janssen}\ \emph {et~al.}(2018)\citenamefont
  {Janssen}, \citenamefont {Chapuis},\ and\ \citenamefont
  {Boissieu}}]{quasicrystalbook}%
  \BibitemOpen
  \bibfield  {author} {\bibinfo {author} {\bibfnamefont {T.}~\bibnamefont
  {Janssen}}, \bibinfo {author} {\bibfnamefont {G.}~\bibnamefont {Chapuis}}, \
  and\ \bibinfo {author} {\bibfnamefont {M.}~\bibnamefont {Boissieu}},\
  }\href@noop {} {\emph {\bibinfo {title} {Aperiodic Crystals: From Modulated
  Phases to Quasicrystals}}}\ (\bibinfo  {publisher} {Oxford Univ. Press},\
  \bibinfo {year} {2018})\BibitemShut {NoStop}%
\bibitem [{\citenamefont {Steurer}(2004)}]{quasicrystalreview2004}%
  \BibitemOpen
  \bibfield  {author} {\bibinfo {author} {\bibfnamefont {W.}~\bibnamefont
  {Steurer}},\ }\href {\doibase 10.1524/zkri.219.7.391.35643} {\bibfield
  {journal} {\bibinfo  {journal} {Z. Kristallogr.}\ }\textbf {\bibinfo {volume}
  {219}},\ \bibinfo {pages} {391} (\bibinfo {year} {2004})}\BibitemShut
  {NoStop}%
\bibitem [{\citenamefont {Lubin}\ \emph {et~al.}(2013)\citenamefont {Lubin},
  \citenamefont {Hryn}, \citenamefont {Huntington} \emph
  {et~al.}}]{plasmon2013}%
  \BibitemOpen
  \bibfield  {author} {\bibinfo {author} {\bibfnamefont {S.~M.}\ \bibnamefont
  {Lubin}}, \bibinfo {author} {\bibfnamefont {A.~J.}\ \bibnamefont {Hryn}},
  \bibinfo {author} {\bibfnamefont {M.~D.}\ \bibnamefont {Huntington}},  \emph
  {et~al.},\ }\href {\doibase 10.1021/nn404703z} {\bibfield  {journal}
  {\bibinfo  {journal} {ACS Nano}\ }\textbf {\bibinfo {volume} {7}},\ \bibinfo
  {pages} {11035} (\bibinfo {year} {2013})}\BibitemShut {NoStop}%
\bibitem [{\citenamefont {Ahn}\ \emph {et~al.}(2018)\citenamefont {Ahn},
  \citenamefont {Moon}, \citenamefont {Kim} \emph
  {et~al.}}]{graphenequasicrystal2018}%
  \BibitemOpen
  \bibfield  {author} {\bibinfo {author} {\bibfnamefont {S.~J.}\ \bibnamefont
  {Ahn}}, \bibinfo {author} {\bibfnamefont {P.}~\bibnamefont {Moon}}, \bibinfo
  {author} {\bibfnamefont {T.-H.}\ \bibnamefont {Kim}},  \emph {et~al.},\
  }\href {\doibase 10.1126/science.aar8412} {\bibfield  {journal} {\bibinfo
  {journal} {Science}\ }\textbf {\bibinfo {volume} {361}},\ \bibinfo {pages}
  {782} (\bibinfo {year} {2018})}\BibitemShut {NoStop}%
\bibitem [{\citenamefont {Viebahn}\ \emph {et~al.}(2019)\citenamefont
  {Viebahn}, \citenamefont {Sbroscia}, \citenamefont {Carter} \emph
  {et~al.}}]{matter2019}%
  \BibitemOpen
  \bibfield  {author} {\bibinfo {author} {\bibfnamefont {K.}~\bibnamefont
  {Viebahn}}, \bibinfo {author} {\bibfnamefont {M.}~\bibnamefont {Sbroscia}},
  \bibinfo {author} {\bibfnamefont {E.}~\bibnamefont {Carter}},  \emph
  {et~al.},\ }\href {\doibase 10.1103/PhysRevLett.122.110404} {\bibfield
  {journal} {\bibinfo  {journal} {Phys. Rev. Lett.}\ }\textbf {\bibinfo
  {volume} {122}},\ \bibinfo {pages} {110404} (\bibinfo {year}
  {2019})}\BibitemShut {NoStop}%
\bibitem [{\citenamefont {Wang}\ \emph {et~al.}(2020)\citenamefont {Wang},
  \citenamefont {Zheng}, \citenamefont {Chen} \emph {et~al.}}]{Ye2020}%
  \BibitemOpen
  \bibfield  {author} {\bibinfo {author} {\bibfnamefont {P.}~\bibnamefont
  {Wang}}, \bibinfo {author} {\bibfnamefont {Y.}~\bibnamefont {Zheng}},
  \bibinfo {author} {\bibfnamefont {X.}~\bibnamefont {Chen}},  \emph {et~al.},\
  }\href {\doibase 10.1038/s41586-019-1851-6} {\bibfield  {journal} {\bibinfo
  {journal} {Nature}\ }\textbf {\bibinfo {volume} {577}},\ \bibinfo {pages} {1}
  (\bibinfo {year} {2020})}\BibitemShut {NoStop}%
\bibitem [{\citenamefont {Yao}\ \emph {et~al.}(2018)\citenamefont {Yao},
  \citenamefont {Wang}, \citenamefont {Bao} \emph {et~al.}}]{zhoushuyun2018}%
  \BibitemOpen
  \bibfield  {author} {\bibinfo {author} {\bibfnamefont {W.}~\bibnamefont
  {Yao}}, \bibinfo {author} {\bibfnamefont {E.}~\bibnamefont {Wang}}, \bibinfo
  {author} {\bibfnamefont {C.}~\bibnamefont {Bao}},  \emph {et~al.},\ }\href
  {\doibase 10.1073/pnas.1720865115} {\bibfield  {journal} {\bibinfo  {journal}
  {Proc. Natl. Acad. Sci.}\ }\textbf {\bibinfo {volume} {115}},\ \bibinfo
  {pages} {6928} (\bibinfo {year} {2018})}\BibitemShut {NoStop}%
\bibitem [{\citenamefont {Diener}\ \emph {et~al.}(2001)\citenamefont {Diener},
  \citenamefont {Georgakis}, \citenamefont {Zhong} \emph {et~al.}}]{niu2001}%
  \BibitemOpen
  \bibfield  {author} {\bibinfo {author} {\bibfnamefont {R.~B.}\ \bibnamefont
  {Diener}}, \bibinfo {author} {\bibfnamefont {G.~A.}\ \bibnamefont
  {Georgakis}}, \bibinfo {author} {\bibfnamefont {J.}~\bibnamefont {Zhong}},
  \emph {et~al.},\ }\href {\doibase 10.1103/PhysRevA.64.033416} {\bibfield
  {journal} {\bibinfo  {journal} {Phys. Rev. A}\ }\textbf {\bibinfo {volume}
  {64}},\ \bibinfo {pages} {033416} (\bibinfo {year} {2001})}\BibitemShut
  {NoStop}%
\bibitem [{\citenamefont {Das~Sarma}\ \emph {et~al.}(1988)\citenamefont
  {Das~Sarma}, \citenamefont {He},\ and\ \citenamefont {Xie}}]{xie1988}%
  \BibitemOpen
  \bibfield  {author} {\bibinfo {author} {\bibfnamefont {S.}~\bibnamefont
  {Das~Sarma}}, \bibinfo {author} {\bibfnamefont {S.}~\bibnamefont {He}}, \
  and\ \bibinfo {author} {\bibfnamefont {X.~C.}\ \bibnamefont {Xie}},\ }\href
  {\doibase 10.1103/PhysRevLett.61.2144} {\bibfield  {journal} {\bibinfo
  {journal} {Phys. Rev. Lett.}\ }\textbf {\bibinfo {volume} {61}},\ \bibinfo
  {pages} {2144} (\bibinfo {year} {1988})}\BibitemShut {NoStop}%
\bibitem [{\citenamefont {Boers}\ \emph {et~al.}(2007)\citenamefont {Boers},
  \citenamefont {Goedeke}, \citenamefont {Hinrichs} \emph
  {et~al.}}]{boers2007}%
  \BibitemOpen
  \bibfield  {author} {\bibinfo {author} {\bibfnamefont {D.~J.}\ \bibnamefont
  {Boers}}, \bibinfo {author} {\bibfnamefont {B.}~\bibnamefont {Goedeke}},
  \bibinfo {author} {\bibfnamefont {D.}~\bibnamefont {Hinrichs}},  \emph
  {et~al.},\ }\href {\doibase 10.1103/PhysRevA.75.063404} {\bibfield  {journal}
  {\bibinfo  {journal} {Phys. Rev. A}\ }\textbf {\bibinfo {volume} {75}},\
  \bibinfo {pages} {063404} (\bibinfo {year} {2007})}\BibitemShut {NoStop}%
\bibitem [{\citenamefont {Li}\ \emph {et~al.}(2017)\citenamefont {Li},
  \citenamefont {Li},\ and\ \citenamefont {Das~Sarma}}]{lixiao2017}%
  \BibitemOpen
  \bibfield  {author} {\bibinfo {author} {\bibfnamefont {X.}~\bibnamefont
  {Li}}, \bibinfo {author} {\bibfnamefont {X.}~\bibnamefont {Li}}, \ and\
  \bibinfo {author} {\bibfnamefont {S.}~\bibnamefont {Das~Sarma}},\ }\href
  {\doibase 10.1103/PhysRevB.96.085119} {\bibfield  {journal} {\bibinfo
  {journal} {Phys. Rev. B}\ }\textbf {\bibinfo {volume} {96}},\ \bibinfo
  {pages} {085119} (\bibinfo {year} {2017})}\BibitemShut {NoStop}%
\bibitem [{\citenamefont {L\"uschen}\ \emph {et~al.}(2018)\citenamefont
  {L\"uschen}, \citenamefont {Scherg}, \citenamefont {Kohlert} \emph
  {et~al.}}]{bloch2018}%
  \BibitemOpen
  \bibfield  {author} {\bibinfo {author} {\bibfnamefont {H.~P.}\ \bibnamefont
  {L\"uschen}}, \bibinfo {author} {\bibfnamefont {S.}~\bibnamefont {Scherg}},
  \bibinfo {author} {\bibfnamefont {T.}~\bibnamefont {Kohlert}},  \emph
  {et~al.},\ }\href {\doibase 10.1103/PhysRevLett.120.160404} {\bibfield
  {journal} {\bibinfo  {journal} {Phys. Rev. Lett.}\ }\textbf {\bibinfo
  {volume} {120}},\ \bibinfo {pages} {160404} (\bibinfo {year}
  {2018})}\BibitemShut {NoStop}%
\bibitem [{\citenamefont {Yao}\ \emph {et~al.}(2019)\citenamefont {Yao},
  \citenamefont {Khoudli}, \citenamefont {Bresque} \emph
  {et~al.}}]{yaohepeng2019}%
  \BibitemOpen
  \bibfield  {author} {\bibinfo {author} {\bibfnamefont {H.}~\bibnamefont
  {Yao}}, \bibinfo {author} {\bibfnamefont {A.}~\bibnamefont {Khoudli}},
  \bibinfo {author} {\bibfnamefont {L.}~\bibnamefont {Bresque}},  \emph
  {et~al.},\ }\href {\doibase 10.1103/PhysRevLett.123.070405} {\bibfield
  {journal} {\bibinfo  {journal} {Phys. Rev. Lett.}\ }\textbf {\bibinfo
  {volume} {123}},\ \bibinfo {pages} {070405} (\bibinfo {year}
  {2019})}\BibitemShut {NoStop}%
\bibitem [{\citenamefont {Wang}\ \emph {et~al.}(2022)\citenamefont {Wang},
  \citenamefont {Zhang}, \citenamefont {Li} \emph {et~al.}}]{wangyunfei2022}%
  \BibitemOpen
  \bibfield  {author} {\bibinfo {author} {\bibfnamefont {Y.}~\bibnamefont
  {Wang}}, \bibinfo {author} {\bibfnamefont {J.-H.}\ \bibnamefont {Zhang}},
  \bibinfo {author} {\bibfnamefont {Y.}~\bibnamefont {Li}},  \emph {et~al.},\
  }\href {\doibase 10.1103/PhysRevLett.129.103401} {\bibfield  {journal}
  {\bibinfo  {journal} {Phys. Rev. Lett.}\ }\textbf {\bibinfo {volume} {129}},\
  \bibinfo {pages} {103401} (\bibinfo {year} {2022})}\BibitemShut {NoStop}%
\bibitem [{\citenamefont {Xiao}\ \emph {et~al.}(2021)\citenamefont {Xiao},
  \citenamefont {Xie}, \citenamefont {Dong} \emph {et~al.}}]{xiaoteng2021}%
  \BibitemOpen
  \bibfield  {author} {\bibinfo {author} {\bibfnamefont {T.}~\bibnamefont
  {Xiao}}, \bibinfo {author} {\bibfnamefont {D.}~\bibnamefont {Xie}}, \bibinfo
  {author} {\bibfnamefont {Z.}~\bibnamefont {Dong}},  \emph {et~al.},\ }\href
  {\doibase https://doi.org/10.1016/j.scib.2021.07.025} {\bibfield  {journal}
  {\bibinfo  {journal} {Sci. Bull.}\ }\textbf {\bibinfo {volume} {66}},\
  \bibinfo {pages} {2175} (\bibinfo {year} {2021})}\BibitemShut {NoStop}%
\bibitem [{\citenamefont {Kohlert}\ \emph {et~al.}(2019)\citenamefont
  {Kohlert}, \citenamefont {Scherg}, \citenamefont {Li} \emph
  {et~al.}}]{kohlert2019}%
  \BibitemOpen
  \bibfield  {author} {\bibinfo {author} {\bibfnamefont {T.}~\bibnamefont
  {Kohlert}}, \bibinfo {author} {\bibfnamefont {S.}~\bibnamefont {Scherg}},
  \bibinfo {author} {\bibfnamefont {X.}~\bibnamefont {Li}},  \emph {et~al.},\
  }\href {\doibase 10.1103/PhysRevLett.122.170403} {\bibfield  {journal}
  {\bibinfo  {journal} {Phys. Rev. Lett.}\ }\textbf {\bibinfo {volume} {122}},\
  \bibinfo {pages} {170403} (\bibinfo {year} {2019})}\BibitemShut {NoStop}%
\bibitem [{\citenamefont {Park}\ \emph {et~al.}(2019)\citenamefont {Park},
  \citenamefont {Kim},\ and\ \citenamefont {Lee}}]{lee2019}%
  \BibitemOpen
  \bibfield  {author} {\bibinfo {author} {\bibfnamefont {M.~J.}\ \bibnamefont
  {Park}}, \bibinfo {author} {\bibfnamefont {H.~S.}\ \bibnamefont {Kim}}, \
  and\ \bibinfo {author} {\bibfnamefont {S.}~\bibnamefont {Lee}},\ }\href
  {\doibase 10.1103/PhysRevB.99.245401} {\bibfield  {journal} {\bibinfo
  {journal} {Phys. Rev. B}\ }\textbf {\bibinfo {volume} {99}},\ \bibinfo
  {pages} {245401} (\bibinfo {year} {2019})}\BibitemShut {NoStop}%
\bibitem [{\citenamefont {Liu}\ \emph {et~al.}(2017)\citenamefont {Liu},
  \citenamefont {Xianlong}, \citenamefont {Chen} \emph {et~al.}}]{guohao2017}%
  \BibitemOpen
  \bibfield  {author} {\bibinfo {author} {\bibfnamefont {T.}~\bibnamefont
  {Liu}}, \bibinfo {author} {\bibfnamefont {G.}~\bibnamefont {Xianlong}},
  \bibinfo {author} {\bibfnamefont {S.}~\bibnamefont {Chen}},  \emph {et~al.},\
  }\href {\doibase https://doi.org/10.1016/j.physleta.2017.09.033} {\bibfield
  {journal} {\bibinfo  {journal} {Phys. Lett. A}\ }\textbf {\bibinfo {volume}
  {381}},\ \bibinfo {pages} {3683} (\bibinfo {year} {2017})}\BibitemShut
  {NoStop}%
\bibitem [{\citenamefont {Zhang}\ \emph {et~al.}(2023)\citenamefont {Zhang},
  \citenamefont {Xie}, \citenamefont {Wu} \emph {et~al.}}]{Zhang20232}%
  \BibitemOpen
  \bibfield  {author} {\bibinfo {author} {\bibfnamefont {S.}~\bibnamefont
  {Zhang}}, \bibinfo {author} {\bibfnamefont {B.}~\bibnamefont {Xie}}, \bibinfo
  {author} {\bibfnamefont {Q.}~\bibnamefont {Wu}},  \emph {et~al.},\ }\href
  {\doibase 10.1021/acs.nanolett.3c00275} {\bibfield  {journal} {\bibinfo
  {journal} {Nano. Lett.}\ }\textbf {\bibinfo {volume} {23}},\ \bibinfo {pages}
  {2921} (\bibinfo {year} {2023})}\BibitemShut {NoStop}%
\bibitem [{\citenamefont {Wu}\ and\ \citenamefont
  {Das~Sarma}(2020)}]{Wudoublebilayer2}%
  \BibitemOpen
  \bibfield  {author} {\bibinfo {author} {\bibfnamefont {F.}~\bibnamefont
  {Wu}}\ and\ \bibinfo {author} {\bibfnamefont {S.}~\bibnamefont {Das~Sarma}},\
  }\href {\doibase 10.1103/PhysRevB.101.155149} {\bibfield  {journal} {\bibinfo
   {journal} {Phys. Rev. B}\ }\textbf {\bibinfo {volume} {101}},\ \bibinfo
  {pages} {155149} (\bibinfo {year} {2020})}\BibitemShut {NoStop}%
\bibitem [{\citenamefont {Ma}\ \emph {et~al.}(2021)\citenamefont {Ma},
  \citenamefont {Li}, \citenamefont {Zheng} \emph {et~al.}}]{Ma2021}%
  \BibitemOpen
  \bibfield  {author} {\bibinfo {author} {\bibfnamefont {Z.}~\bibnamefont
  {Ma}}, \bibinfo {author} {\bibfnamefont {S.}~\bibnamefont {Li}}, \bibinfo
  {author} {\bibfnamefont {Y.-W.}\ \bibnamefont {Zheng}},  \emph {et~al.},\
  }\href {https://www.sciencedirect.com/science/article/pii/S2095927320306575}
  {\bibfield  {journal} {\bibinfo  {journal} {Sci. Bull.}\ }\textbf {\bibinfo
  {volume} {66}},\ \bibinfo {pages} {18} (\bibinfo {year} {2021})}\BibitemShut
  {NoStop}%
\bibitem [{\citenamefont {Rademaker}\ \emph {et~al.}(2020)\citenamefont
  {Rademaker}, \citenamefont {Protopopov},\ and\ \citenamefont
  {Abanin}}]{rademaker2020prr}%
  \BibitemOpen
  \bibfield  {author} {\bibinfo {author} {\bibfnamefont {L.}~\bibnamefont
  {Rademaker}}, \bibinfo {author} {\bibfnamefont {I.~V.}\ \bibnamefont
  {Protopopov}}, \ and\ \bibinfo {author} {\bibfnamefont {D.~A.}\ \bibnamefont
  {Abanin}},\ }\href {\doibase 10.1103/PhysRevResearch.2.033150} {\bibfield
  {journal} {\bibinfo  {journal} {Phys. Rev. Res.}\ }\textbf {\bibinfo {volume}
  {2}},\ \bibinfo {pages} {033150} (\bibinfo {year} {2020})}\BibitemShut
  {NoStop}%
\bibitem [{\citenamefont {Koshino}(2019)}]{Koshino20191}%
  \BibitemOpen
  \bibfield  {author} {\bibinfo {author} {\bibfnamefont {M.}~\bibnamefont
  {Koshino}},\ }\href {\doibase 10.1103/PhysRevB.99.235406} {\bibfield
  {journal} {\bibinfo  {journal} {Phys. Rev. B}\ }\textbf {\bibinfo {volume}
  {99}},\ \bibinfo {pages} {235406} (\bibinfo {year} {2019})}\BibitemShut
  {NoStop}%
\bibitem [{\citenamefont {Chebrolu}\ \emph {et~al.}(2019)\citenamefont
  {Chebrolu}, \citenamefont {Chittari},\ and\ \citenamefont
  {Jung}}]{jung20191}%
  \BibitemOpen
  \bibfield  {author} {\bibinfo {author} {\bibfnamefont {N.~R.}\ \bibnamefont
  {Chebrolu}}, \bibinfo {author} {\bibfnamefont {B.~L.}\ \bibnamefont
  {Chittari}}, \ and\ \bibinfo {author} {\bibfnamefont {J.}~\bibnamefont
  {Jung}},\ }\href {\doibase 10.1103/PhysRevB.99.235417} {\bibfield  {journal}
  {\bibinfo  {journal} {Phys. Rev. B}\ }\textbf {\bibinfo {volume} {99}},\
  \bibinfo {pages} {235417} (\bibinfo {year} {2019})}\BibitemShut {NoStop}%
\bibitem [{\citenamefont {Liu}\ \emph {et~al.}(2019)\citenamefont {Liu},
  \citenamefont {Ma}, \citenamefont {Gao} \emph {et~al.}}]{PhysRevX.9.031021}%
  \BibitemOpen
  \bibfield  {author} {\bibinfo {author} {\bibfnamefont {J.}~\bibnamefont
  {Liu}}, \bibinfo {author} {\bibfnamefont {Z.}~\bibnamefont {Ma}}, \bibinfo
  {author} {\bibfnamefont {J.}~\bibnamefont {Gao}},  \emph {et~al.},\ }\href
  {\doibase 10.1103/PhysRevX.9.031021} {\bibfield  {journal} {\bibinfo
  {journal} {Phys. Rev. X}\ }\textbf {\bibinfo {volume} {9}},\ \bibinfo {pages}
  {031021} (\bibinfo {year} {2019})}\BibitemShut {NoStop}%
\bibitem [{\citenamefont {Haddadi}\ \emph {et~al.}(2020)\citenamefont
  {Haddadi}, \citenamefont {Wu}, \citenamefont {Kruchkov} \emph
  {et~al.}}]{doi:10.1021/acs.nanolett.9b05117}%
  \BibitemOpen
  \bibfield  {author} {\bibinfo {author} {\bibfnamefont {F.}~\bibnamefont
  {Haddadi}}, \bibinfo {author} {\bibfnamefont {Q.}~\bibnamefont {Wu}},
  \bibinfo {author} {\bibfnamefont {A.~J.}\ \bibnamefont {Kruchkov}},  \emph
  {et~al.},\ }\href {\doibase 10.1021/acs.nanolett.9b05117} {\bibfield
  {journal} {\bibinfo  {journal} {Nano Lett.}\ }\textbf {\bibinfo {volume}
  {20}},\ \bibinfo {pages} {2410} (\bibinfo {year} {2020})}\BibitemShut
  {NoStop}%
\bibitem [{\citenamefont {Park}\ \emph {et~al.}(2020)\citenamefont {Park},
  \citenamefont {Chittari},\ and\ \citenamefont {Jung}}]{jj2020prb}%
  \BibitemOpen
  \bibfield  {author} {\bibinfo {author} {\bibfnamefont {Y.}~\bibnamefont
  {Park}}, \bibinfo {author} {\bibfnamefont {B.~L.}\ \bibnamefont {Chittari}},
  \ and\ \bibinfo {author} {\bibfnamefont {J.}~\bibnamefont {Jung}},\ }\href
  {\doibase 10.1103/PhysRevB.102.035411} {\bibfield  {journal} {\bibinfo
  {journal} {Phys. Rev. B}\ }\textbf {\bibinfo {volume} {102}},\ \bibinfo
  {pages} {035411} (\bibinfo {year} {2020})}\BibitemShut {NoStop}%
\bibitem [{\citenamefont {Ma}\ \emph {et~al.}(2022{\natexlab{a}})\citenamefont
  {Ma}, \citenamefont {Li}, \citenamefont {Xiao} \emph {et~al.}}]{mazhen2023}%
  \BibitemOpen
  \bibfield  {author} {\bibinfo {author} {\bibfnamefont {Z.}~\bibnamefont
  {Ma}}, \bibinfo {author} {\bibfnamefont {S.}~\bibnamefont {Li}}, \bibinfo
  {author} {\bibfnamefont {M.-M.}\ \bibnamefont {Xiao}},  \emph {et~al.},\
  }\href {\doibase 10.1007/s11467-022-1220-z} {\bibfield  {journal} {\bibinfo
  {journal} {Front. Phys.}\ }\textbf {\bibinfo {volume} {18}},\ \bibinfo
  {pages} {13307} (\bibinfo {year} {2022}{\natexlab{a}})}\BibitemShut {NoStop}%
\bibitem [{\citenamefont {Wu}\ \emph {et~al.}(2018)\citenamefont {Wu},
  \citenamefont {Lovorn}, \citenamefont {Tutuc} \emph {et~al.}}]{wu20182}%
  \BibitemOpen
  \bibfield  {author} {\bibinfo {author} {\bibfnamefont {F.}~\bibnamefont
  {Wu}}, \bibinfo {author} {\bibfnamefont {T.}~\bibnamefont {Lovorn}}, \bibinfo
  {author} {\bibfnamefont {E.}~\bibnamefont {Tutuc}},  \emph {et~al.},\ }\href
  {\doibase 10.1103/PhysRevLett.121.026402} {\bibfield  {journal} {\bibinfo
  {journal} {Phys. Rev. Lett.}\ }\textbf {\bibinfo {volume} {121}},\ \bibinfo
  {pages} {026402} (\bibinfo {year} {2018})}\BibitemShut {NoStop}%
\bibitem [{\citenamefont {Tarnopolsky}\ \emph {et~al.}(2019)\citenamefont
  {Tarnopolsky}, \citenamefont {Kruchkov},\ and\ \citenamefont
  {Vishwanath}}]{ashvin2019}%
  \BibitemOpen
  \bibfield  {author} {\bibinfo {author} {\bibfnamefont {G.}~\bibnamefont
  {Tarnopolsky}}, \bibinfo {author} {\bibfnamefont {A.~J.}\ \bibnamefont
  {Kruchkov}}, \ and\ \bibinfo {author} {\bibfnamefont {A.}~\bibnamefont
  {Vishwanath}},\ }\href {\doibase 10.1103/PhysRevLett.122.106405} {\bibfield
  {journal} {\bibinfo  {journal} {Phys. Rev. Lett.}\ }\textbf {\bibinfo
  {volume} {122}},\ \bibinfo {pages} {106405} (\bibinfo {year}
  {2019})}\BibitemShut {NoStop}%
\bibitem [{\citenamefont {Guo}\ \emph {et~al.}(2018)\citenamefont {Guo},
  \citenamefont {Zhu}, \citenamefont {Feng},\ and\ \citenamefont
  {Scalettar}}]{guohuaiming2018}%
  \BibitemOpen
  \bibfield  {author} {\bibinfo {author} {\bibfnamefont {H.}~\bibnamefont
  {Guo}}, \bibinfo {author} {\bibfnamefont {X.}~\bibnamefont {Zhu}}, \bibinfo
  {author} {\bibfnamefont {S.}~\bibnamefont {Feng}}, \ and\ \bibinfo {author}
  {\bibfnamefont {R.~T.}\ \bibnamefont {Scalettar}},\ }\href {\doibase
  10.1103/PhysRevB.97.235453} {\bibfield  {journal} {\bibinfo  {journal} {Phys.
  Rev. B}\ }\textbf {\bibinfo {volume} {97}},\ \bibinfo {pages} {235453}
  (\bibinfo {year} {2018})}\BibitemShut {NoStop}%
\bibitem [{\citenamefont {Kennes}\ \emph {et~al.}(2018)\citenamefont {Kennes},
  \citenamefont {Lischner},\ and\ \citenamefont {Karrasch}}]{kennes2018}%
  \BibitemOpen
  \bibfield  {author} {\bibinfo {author} {\bibfnamefont {D.~M.}\ \bibnamefont
  {Kennes}}, \bibinfo {author} {\bibfnamefont {J.}~\bibnamefont {Lischner}}, \
  and\ \bibinfo {author} {\bibfnamefont {C.}~\bibnamefont {Karrasch}},\ }\href
  {\doibase 10.1103/PhysRevB.98.241407} {\bibfield  {journal} {\bibinfo
  {journal} {Phys. Rev. B}\ }\textbf {\bibinfo {volume} {98}},\ \bibinfo
  {pages} {241407} (\bibinfo {year} {2018})}\BibitemShut {NoStop}%
\bibitem [{\citenamefont {Peltonen}\ \emph {et~al.}(2018)\citenamefont
  {Peltonen}, \citenamefont {Ojaj\"arvi},\ and\ \citenamefont
  {Heikkil\"a}}]{meanfield2018}%
  \BibitemOpen
  \bibfield  {author} {\bibinfo {author} {\bibfnamefont {T.~J.}\ \bibnamefont
  {Peltonen}}, \bibinfo {author} {\bibfnamefont {R.}~\bibnamefont
  {Ojaj\"arvi}}, \ and\ \bibinfo {author} {\bibfnamefont {T.~T.}\ \bibnamefont
  {Heikkil\"a}},\ }\href {\doibase 10.1103/PhysRevB.98.220504} {\bibfield
  {journal} {\bibinfo  {journal} {Phys. Rev. B}\ }\textbf {\bibinfo {volume}
  {98}},\ \bibinfo {pages} {220504} (\bibinfo {year} {2018})}\BibitemShut
  {NoStop}%
\bibitem [{\citenamefont {Xie}\ \emph {et~al.}(2020)\citenamefont {Xie},
  \citenamefont {Song}, \citenamefont {Lian} \emph {et~al.}}]{xiefang2020}%
  \BibitemOpen
  \bibfield  {author} {\bibinfo {author} {\bibfnamefont {F.}~\bibnamefont
  {Xie}}, \bibinfo {author} {\bibfnamefont {Z.}~\bibnamefont {Song}}, \bibinfo
  {author} {\bibfnamefont {B.}~\bibnamefont {Lian}},  \emph {et~al.},\ }\href
  {\doibase 10.1103/PhysRevLett.124.167002} {\bibfield  {journal} {\bibinfo
  {journal} {Phys. Rev. Lett.}\ }\textbf {\bibinfo {volume} {124}},\ \bibinfo
  {pages} {167002} (\bibinfo {year} {2020})}\BibitemShut {NoStop}%
\bibitem [{\citenamefont {Julku}\ \emph {et~al.}(2020)\citenamefont {Julku},
  \citenamefont {Peltonen}, \citenamefont {Liang} \emph {et~al.}}]{torma2020}%
  \BibitemOpen
  \bibfield  {author} {\bibinfo {author} {\bibfnamefont {A.}~\bibnamefont
  {Julku}}, \bibinfo {author} {\bibfnamefont {T.~J.}\ \bibnamefont {Peltonen}},
  \bibinfo {author} {\bibfnamefont {L.}~\bibnamefont {Liang}},  \emph
  {et~al.},\ }\href {\doibase 10.1103/PhysRevB.101.060505} {\bibfield
  {journal} {\bibinfo  {journal} {Phys. Rev. B}\ }\textbf {\bibinfo {volume}
  {101}},\ \bibinfo {pages} {060505} (\bibinfo {year} {2020})}\BibitemShut
  {NoStop}%
\bibitem [{\citenamefont {Gonz\'alez}\ and\ \citenamefont
  {Stauber}(2019)}]{kohn2019}%
  \BibitemOpen
  \bibfield  {author} {\bibinfo {author} {\bibfnamefont {J.}~\bibnamefont
  {Gonz\'alez}}\ and\ \bibinfo {author} {\bibfnamefont {T.}~\bibnamefont
  {Stauber}},\ }\href {\doibase 10.1103/PhysRevLett.122.026801} {\bibfield
  {journal} {\bibinfo  {journal} {Phys. Rev. Lett.}\ }\textbf {\bibinfo
  {volume} {122}},\ \bibinfo {pages} {026801} (\bibinfo {year}
  {2019})}\BibitemShut {NoStop}%
\bibitem [{\citenamefont {Wu}\ \emph {et~al.}(2020)\citenamefont {Wu},
  \citenamefont {Hanke}, \citenamefont {Fink} \emph {et~al.}}]{wuxianxin2020}%
  \BibitemOpen
  \bibfield  {author} {\bibinfo {author} {\bibfnamefont {X.}~\bibnamefont
  {Wu}}, \bibinfo {author} {\bibfnamefont {W.}~\bibnamefont {Hanke}}, \bibinfo
  {author} {\bibfnamefont {M.}~\bibnamefont {Fink}},  \emph {et~al.},\ }\href
  {\doibase 10.1103/PhysRevB.101.134517} {\bibfield  {journal} {\bibinfo
  {journal} {Phys. Rev. B}\ }\textbf {\bibinfo {volume} {101}},\ \bibinfo
  {pages} {134517} (\bibinfo {year} {2020})}\BibitemShut {NoStop}%
\bibitem [{\citenamefont {Lian}\ \emph {et~al.}(2019)\citenamefont {Lian},
  \citenamefont {Wang},\ and\ \citenamefont {Bernevig}}]{lianbiao2019}%
  \BibitemOpen
  \bibfield  {author} {\bibinfo {author} {\bibfnamefont {B.}~\bibnamefont
  {Lian}}, \bibinfo {author} {\bibfnamefont {Z.}~\bibnamefont {Wang}}, \ and\
  \bibinfo {author} {\bibfnamefont {B.~A.}\ \bibnamefont {Bernevig}},\ }\href
  {\doibase 10.1103/PhysRevLett.122.257002} {\bibfield  {journal} {\bibinfo
  {journal} {Phys. Rev. Lett.}\ }\textbf {\bibinfo {volume} {122}},\ \bibinfo
  {pages} {257002} (\bibinfo {year} {2019})}\BibitemShut {NoStop}%
\bibitem [{\citenamefont {Tang}\ \emph {et~al.}(2019)\citenamefont {Tang},
  \citenamefont {Yang}, \citenamefont {Wang} \emph
  {et~al.}}]{wangqianghua2019}%
  \BibitemOpen
  \bibfield  {author} {\bibinfo {author} {\bibfnamefont {Q.-K.}\ \bibnamefont
  {Tang}}, \bibinfo {author} {\bibfnamefont {L.}~\bibnamefont {Yang}}, \bibinfo
  {author} {\bibfnamefont {D.}~\bibnamefont {Wang}},  \emph {et~al.},\ }\href
  {\doibase 10.1103/PhysRevB.99.094521} {\bibfield  {journal} {\bibinfo
  {journal} {Phys. Rev. B}\ }\textbf {\bibinfo {volume} {99}},\ \bibinfo
  {pages} {094521} (\bibinfo {year} {2019})}\BibitemShut {NoStop}%
\bibitem [{\citenamefont {You}\ and\ \citenamefont
  {Vishwanath}(2019)}]{youyizhuang2019}%
  \BibitemOpen
  \bibfield  {author} {\bibinfo {author} {\bibfnamefont {Y.-Z.}\ \bibnamefont
  {You}}\ and\ \bibinfo {author} {\bibfnamefont {A.}~\bibnamefont
  {Vishwanath}},\ }\href {\doibase 10.1038/s41535-019-0153-4} {\bibfield
  {journal} {\bibinfo  {journal} {npj Quantum Mater.}\ }\textbf {\bibinfo
  {volume} {4}},\ \bibinfo {pages} {16} (\bibinfo {year} {2019})}\BibitemShut
  {NoStop}%
\bibitem [{\citenamefont {Roy}\ and\ \citenamefont {Juri\ifmmode \check{c}\else
  \v{c}\fi{}i\ifmmode~\acute{c}\else \'{c}\fi{}}(2019)}]{roy2019}%
  \BibitemOpen
  \bibfield  {author} {\bibinfo {author} {\bibfnamefont {B.}~\bibnamefont
  {Roy}}\ and\ \bibinfo {author} {\bibfnamefont {V.}~\bibnamefont {Juri\ifmmode
  \check{c}\else \v{c}\fi{}i\ifmmode~\acute{c}\else \'{c}\fi{}}},\ }\href
  {\doibase 10.1103/PhysRevB.99.121407} {\bibfield  {journal} {\bibinfo
  {journal} {Phys. Rev. B}\ }\textbf {\bibinfo {volume} {99}},\ \bibinfo
  {pages} {121407} (\bibinfo {year} {2019})}\BibitemShut {NoStop}%
\bibitem [{\citenamefont {Kang}\ and\ \citenamefont
  {Vafek}(2019)}]{kangjian2019}%
  \BibitemOpen
  \bibfield  {author} {\bibinfo {author} {\bibfnamefont {J.}~\bibnamefont
  {Kang}}\ and\ \bibinfo {author} {\bibfnamefont {O.}~\bibnamefont {Vafek}},\
  }\href {\doibase 10.1103/PhysRevLett.122.246401} {\bibfield  {journal}
  {\bibinfo  {journal} {Phys. Rev. Lett.}\ }\textbf {\bibinfo {volume} {122}},\
  \bibinfo {pages} {246401} (\bibinfo {year} {2019})}\BibitemShut {NoStop}%
\bibitem [{\citenamefont {Zhang}\ \emph {et~al.}(2019)\citenamefont {Zhang},
  \citenamefont {Mao}, \citenamefont {Cao} \emph {et~al.}}]{zhangyahui2019}%
  \BibitemOpen
  \bibfield  {author} {\bibinfo {author} {\bibfnamefont {Y.-H.}\ \bibnamefont
  {Zhang}}, \bibinfo {author} {\bibfnamefont {D.}~\bibnamefont {Mao}}, \bibinfo
  {author} {\bibfnamefont {Y.}~\bibnamefont {Cao}},  \emph {et~al.},\ }\href
  {\doibase 10.1103/PhysRevB.99.075127} {\bibfield  {journal} {\bibinfo
  {journal} {Phys. Rev. B}\ }\textbf {\bibinfo {volume} {99}},\ \bibinfo
  {pages} {075127} (\bibinfo {year} {2019})}\BibitemShut {NoStop}%
\bibitem [{\citenamefont {Liu}\ \emph {et~al.}(2018)\citenamefont {Liu},
  \citenamefont {Zhang}, \citenamefont {Chen} \emph
  {et~al.}}]{liuchengcheng2018}%
  \BibitemOpen
  \bibfield  {author} {\bibinfo {author} {\bibfnamefont {C.-C.}\ \bibnamefont
  {Liu}}, \bibinfo {author} {\bibfnamefont {L.-D.}\ \bibnamefont {Zhang}},
  \bibinfo {author} {\bibfnamefont {W.-Q.}\ \bibnamefont {Chen}},  \emph
  {et~al.},\ }\href {\doibase 10.1103/PhysRevLett.121.217001} {\bibfield
  {journal} {\bibinfo  {journal} {Phys. Rev. Lett.}\ }\textbf {\bibinfo
  {volume} {121}},\ \bibinfo {pages} {217001} (\bibinfo {year}
  {2018})}\BibitemShut {NoStop}%
\bibitem [{\citenamefont {Khalaf}\ \emph {et~al.}(2021)\citenamefont {Khalaf},
  \citenamefont {Chatterjee}, \citenamefont {Bultinck} \emph
  {et~al.}}]{khalaf2021}%
  \BibitemOpen
  \bibfield  {author} {\bibinfo {author} {\bibfnamefont {E.}~\bibnamefont
  {Khalaf}}, \bibinfo {author} {\bibfnamefont {S.}~\bibnamefont {Chatterjee}},
  \bibinfo {author} {\bibfnamefont {N.}~\bibnamefont {Bultinck}},  \emph
  {et~al.},\ }\href {\doibase 10.1126/sciadv.abf5299} {\bibfield  {journal}
  {\bibinfo  {journal} {Sci. Adv.}\ }\textbf {\bibinfo {volume} {7}},\ \bibinfo
  {pages} {eabf5299} (\bibinfo {year} {2021})}\BibitemShut {NoStop}%
\bibitem [{\citenamefont {Zhang}\ \emph {et~al.}(2020)\citenamefont {Zhang},
  \citenamefont {Jiang}, \citenamefont {Wang} \emph {et~al.}}]{zhangyi2020}%
  \BibitemOpen
  \bibfield  {author} {\bibinfo {author} {\bibfnamefont {Y.}~\bibnamefont
  {Zhang}}, \bibinfo {author} {\bibfnamefont {K.}~\bibnamefont {Jiang}},
  \bibinfo {author} {\bibfnamefont {Z.}~\bibnamefont {Wang}},  \emph {et~al.},\
  }\href {\doibase 10.1103/PhysRevB.102.035136} {\bibfield  {journal} {\bibinfo
   {journal} {Phys. Rev. B}\ }\textbf {\bibinfo {volume} {102}},\ \bibinfo
  {pages} {035136} (\bibinfo {year} {2020})}\BibitemShut {NoStop}%
\bibitem [{\citenamefont {Yu}\ \emph {et~al.}(2021)\citenamefont {Yu},
  \citenamefont {Kennes}, \citenamefont {Rubio} \emph {et~al.}}]{yutao2021}%
  \BibitemOpen
  \bibfield  {author} {\bibinfo {author} {\bibfnamefont {T.}~\bibnamefont
  {Yu}}, \bibinfo {author} {\bibfnamefont {D.~M.}\ \bibnamefont {Kennes}},
  \bibinfo {author} {\bibfnamefont {A.}~\bibnamefont {Rubio}},  \emph
  {et~al.},\ }\href {\doibase 10.1103/PhysRevLett.127.127001} {\bibfield
  {journal} {\bibinfo  {journal} {Phys. Rev. Lett.}\ }\textbf {\bibinfo
  {volume} {127}},\ \bibinfo {pages} {127001} (\bibinfo {year}
  {2021})}\BibitemShut {NoStop}%
\bibitem [{\citenamefont {Moon}\ \emph {et~al.}(2019)\citenamefont {Moon},
  \citenamefont {Koshino},\ and\ \citenamefont {Son}}]{moonquasicrystal2019}%
  \BibitemOpen
  \bibfield  {author} {\bibinfo {author} {\bibfnamefont {P.}~\bibnamefont
  {Moon}}, \bibinfo {author} {\bibfnamefont {M.}~\bibnamefont {Koshino}}, \
  and\ \bibinfo {author} {\bibfnamefont {Y.-W.}\ \bibnamefont {Son}},\ }\href
  {\doibase 10.1103/PhysRevB.99.165430} {\bibfield  {journal} {\bibinfo
  {journal} {Phys. Rev. B}\ }\textbf {\bibinfo {volume} {99}},\ \bibinfo
  {pages} {165430} (\bibinfo {year} {2019})}\BibitemShut {NoStop}%
\bibitem [{\citenamefont {Yan}\ \emph {et~al.}(2019)\citenamefont {Yan},
  \citenamefont {Ma}, \citenamefont {Qiao} \emph {et~al.}}]{helin2019}%
  \BibitemOpen
  \bibfield  {author} {\bibinfo {author} {\bibfnamefont {C.}~\bibnamefont
  {Yan}}, \bibinfo {author} {\bibfnamefont {D.-L.}\ \bibnamefont {Ma}},
  \bibinfo {author} {\bibfnamefont {J.-B.}\ \bibnamefont {Qiao}},  \emph
  {et~al.},\ }\href {\doibase 10.1088/2053-1583/ab3b16} {\bibfield  {journal}
  {\bibinfo  {journal} {2D Mater.}\ }\textbf {\bibinfo {volume} {6}},\ \bibinfo
  {pages} {045041} (\bibinfo {year} {2019})}\BibitemShut {NoStop}%
\bibitem [{\citenamefont {Yu}\ \emph {et~al.}(2019)\citenamefont {Yu},
  \citenamefont {Wu}, \citenamefont {Zhan} \emph {et~al.}}]{yuan2019}%
  \BibitemOpen
  \bibfield  {author} {\bibinfo {author} {\bibfnamefont {G.}~\bibnamefont
  {Yu}}, \bibinfo {author} {\bibfnamefont {Z.}~\bibnamefont {Wu}}, \bibinfo
  {author} {\bibfnamefont {Z.}~\bibnamefont {Zhan}},  \emph {et~al.},\ }\href
  {\doibase 10.1038/s41524-019-0258-0} {\bibfield  {journal} {\bibinfo
  {journal} {npj Comput Mater.}\ }\textbf {\bibinfo {volume} {5}},\ \bibinfo
  {pages} {122} (\bibinfo {year} {2019})}\BibitemShut {NoStop}%
\bibitem [{\citenamefont {Suzuki}\ \emph {et~al.}(2019)\citenamefont {Suzuki},
  \citenamefont {Iimori}, \citenamefont {Ahn} \emph {et~al.}}]{suzuki2019}%
  \BibitemOpen
  \bibfield  {author} {\bibinfo {author} {\bibfnamefont {T.}~\bibnamefont
  {Suzuki}}, \bibinfo {author} {\bibfnamefont {T.}~\bibnamefont {Iimori}},
  \bibinfo {author} {\bibfnamefont {S.~J.}\ \bibnamefont {Ahn}},  \emph
  {et~al.},\ }\href {\doibase 10.1021/acsnano.9b06091} {\bibfield  {journal}
  {\bibinfo  {journal} {ACS Nano}\ }\textbf {\bibinfo {volume} {13}},\ \bibinfo
  {pages} {11981} (\bibinfo {year} {2019})}\BibitemShut {NoStop}%
\bibitem [{\citenamefont {Spurrier}\ and\ \citenamefont
  {Cooper}(2019)}]{spur2019}%
  \BibitemOpen
  \bibfield  {author} {\bibinfo {author} {\bibfnamefont {S.}~\bibnamefont
  {Spurrier}}\ and\ \bibinfo {author} {\bibfnamefont {N.~R.}\ \bibnamefont
  {Cooper}},\ }\href {\doibase 10.1103/PhysRevB.100.081405} {\bibfield
  {journal} {\bibinfo  {journal} {Phys. Rev. B}\ }\textbf {\bibinfo {volume}
  {100}},\ \bibinfo {pages} {081405} (\bibinfo {year} {2019})}\BibitemShut
  {NoStop}%
\bibitem [{\citenamefont {Liu}\ \emph {et~al.}(2023)\citenamefont {Liu},
  \citenamefont {Zhang}, \citenamefont {Chen} \emph {et~al.}}]{yangfan2023}%
  \BibitemOpen
  \bibfield  {author} {\bibinfo {author} {\bibfnamefont {Y.-B.}\ \bibnamefont
  {Liu}}, \bibinfo {author} {\bibfnamefont {Y.}~\bibnamefont {Zhang}}, \bibinfo
  {author} {\bibfnamefont {W.-Q.}\ \bibnamefont {Chen}},  \emph {et~al.},\
  }\href {\doibase 10.1103/PhysRevB.107.014501} {\bibfield  {journal} {\bibinfo
   {journal} {Phys. Rev. B}\ }\textbf {\bibinfo {volume} {107}},\ \bibinfo
  {pages} {014501} (\bibinfo {year} {2023})}\BibitemShut {NoStop}%
\bibitem [{\citenamefont {Park}\ \emph {et~al.}(2021)\citenamefont {Park},
  \citenamefont {Cao}, \citenamefont {Watanabe} \emph {et~al.}}]{Park20213}%
  \BibitemOpen
  \bibfield  {author} {\bibinfo {author} {\bibfnamefont {J.~M.}\ \bibnamefont
  {Park}}, \bibinfo {author} {\bibfnamefont {Y.}~\bibnamefont {Cao}}, \bibinfo
  {author} {\bibfnamefont {K.}~\bibnamefont {Watanabe}},  \emph {et~al.},\
  }\href {\doibase 10.1038/s41586-021-03192-0} {\bibfield  {journal} {\bibinfo
  {journal} {Nature}\ }\textbf {\bibinfo {volume} {590}},\ \bibinfo {pages}
  {249} (\bibinfo {year} {2021})}\BibitemShut {NoStop}%
\bibitem [{\citenamefont {Li}\ \emph {et~al.}(2022{\natexlab{a}})\citenamefont
  {Li}, \citenamefont {Xue}, \citenamefont {Fan} \emph {et~al.}}]{Carr20203}%
  \BibitemOpen
  \bibfield  {author} {\bibinfo {author} {\bibfnamefont {Y.}~\bibnamefont
  {Li}}, \bibinfo {author} {\bibfnamefont {M.}~\bibnamefont {Xue}}, \bibinfo
  {author} {\bibfnamefont {H.}~\bibnamefont {Fan}},  \emph {et~al.},\ }\href
  {\doibase 10.1021/acs.nanolett.2c01710} {\bibfield  {journal} {\bibinfo
  {journal} {Nano. Lett.}\ }\textbf {\bibinfo {volume} {22}},\ \bibinfo {pages}
  {6215} (\bibinfo {year} {2022}{\natexlab{a}})}\BibitemShut {NoStop}%
\bibitem [{\citenamefont {Mora}\ \emph {et~al.}(2019)\citenamefont {Mora},
  \citenamefont {Regnault},\ and\ \citenamefont {Bernevig}}]{mora20193}%
  \BibitemOpen
  \bibfield  {author} {\bibinfo {author} {\bibfnamefont {C.}~\bibnamefont
  {Mora}}, \bibinfo {author} {\bibfnamefont {N.}~\bibnamefont {Regnault}}, \
  and\ \bibinfo {author} {\bibfnamefont {B.~A.}\ \bibnamefont {Bernevig}},\
  }\href {\doibase 10.1103/PhysRevLett.123.026402} {\bibfield  {journal}
  {\bibinfo  {journal} {Phys. Rev. Lett.}\ }\textbf {\bibinfo {volume} {123}},\
  \bibinfo {pages} {026402} (\bibinfo {year} {2019})}\BibitemShut {NoStop}%
\bibitem [{\citenamefont {Zhu}\ \emph {et~al.}(2020)\citenamefont {Zhu},
  \citenamefont {Carr}, \citenamefont {Massatt} \emph {et~al.}}]{zhuziyan2020}%
  \BibitemOpen
  \bibfield  {author} {\bibinfo {author} {\bibfnamefont {Z.}~\bibnamefont
  {Zhu}}, \bibinfo {author} {\bibfnamefont {S.}~\bibnamefont {Carr}}, \bibinfo
  {author} {\bibfnamefont {D.}~\bibnamefont {Massatt}},  \emph {et~al.},\
  }\href {\doibase 10.1103/PhysRevLett.125.116404} {\bibfield  {journal}
  {\bibinfo  {journal} {Phys. Rev. Lett.}\ }\textbf {\bibinfo {volume} {125}},\
  \bibinfo {pages} {116404} (\bibinfo {year} {2020})}\BibitemShut {NoStop}%
\bibitem [{\citenamefont {Zuo}\ \emph {et~al.}(2018)\citenamefont {Zuo},
  \citenamefont {Qiao}, \citenamefont {Ma} \emph {et~al.}}]{zuo20183}%
  \BibitemOpen
  \bibfield  {author} {\bibinfo {author} {\bibfnamefont {W.-J.}\ \bibnamefont
  {Zuo}}, \bibinfo {author} {\bibfnamefont {J.-B.}\ \bibnamefont {Qiao}},
  \bibinfo {author} {\bibfnamefont {D.-L.}\ \bibnamefont {Ma}},  \emph
  {et~al.},\ }\href {\doibase 10.1103/PhysRevB.97.035440} {\bibfield  {journal}
  {\bibinfo  {journal} {Phys. Rev. B}\ }\textbf {\bibinfo {volume} {97}},\
  \bibinfo {pages} {035440} (\bibinfo {year} {2018})}\BibitemShut {NoStop}%
\bibitem [{\citenamefont {Zhang}\ \emph {et~al.}(2021)\citenamefont {Zhang},
  \citenamefont {Tsai}, \citenamefont {Zhu} \emph {et~al.}}]{wangke2021}%
  \BibitemOpen
  \bibfield  {author} {\bibinfo {author} {\bibfnamefont {X.}~\bibnamefont
  {Zhang}}, \bibinfo {author} {\bibfnamefont {K.-T.}\ \bibnamefont {Tsai}},
  \bibinfo {author} {\bibfnamefont {Z.}~\bibnamefont {Zhu}},  \emph {et~al.},\
  }\href {\doibase 10.1103/PhysRevLett.127.166802} {\bibfield  {journal}
  {\bibinfo  {journal} {Phys. Rev. Lett.}\ }\textbf {\bibinfo {volume} {127}},\
  \bibinfo {pages} {166802} (\bibinfo {year} {2021})}\BibitemShut {NoStop}%
\bibitem [{\citenamefont {Li}\ \emph {et~al.}(2022{\natexlab{b}})\citenamefont
  {Li}, \citenamefont {Xue}, \citenamefont {Fan} \emph {et~al.}}]{liyuhao2022}%
  \BibitemOpen
  \bibfield  {author} {\bibinfo {author} {\bibfnamefont {Y.}~\bibnamefont
  {Li}}, \bibinfo {author} {\bibfnamefont {M.}~\bibnamefont {Xue}}, \bibinfo
  {author} {\bibfnamefont {H.}~\bibnamefont {Fan}},  \emph {et~al.},\ }\href
  {\doibase 10.1021/acs.nanolett.2c01710} {\bibfield  {journal} {\bibinfo
  {journal} {Nano. Lett.}\ } (\bibinfo {year} {2022}{\natexlab{b}}),\
  10.1021/acs.nanolett.2c01710}\BibitemShut {NoStop}%
\bibitem [{\citenamefont {Liang}\ \emph {et~al.}(2022)\citenamefont {Liang},
  \citenamefont {Xiao}, \citenamefont {Ma} \emph {et~al.}}]{liangmiao2022}%
  \BibitemOpen
  \bibfield  {author} {\bibinfo {author} {\bibfnamefont {M.}~\bibnamefont
  {Liang}}, \bibinfo {author} {\bibfnamefont {M.-M.}\ \bibnamefont {Xiao}},
  \bibinfo {author} {\bibfnamefont {Z.}~\bibnamefont {Ma}},  \emph {et~al.},\
  }\href {\doibase 10.1103/PhysRevB.105.195422} {\bibfield  {journal} {\bibinfo
   {journal} {Phys. Rev. B}\ }\textbf {\bibinfo {volume} {105}},\ \bibinfo
  {pages} {195422} (\bibinfo {year} {2022})}\BibitemShut {NoStop}%
\bibitem [{\citenamefont {Anđelković}\ \emph {et~al.}(2020)\citenamefont
  {Anđelković}, \citenamefont {Milovanovic}, \citenamefont {Covaci} \emph
  {et~al.}}]{peeters2020}%
  \BibitemOpen
  \bibfield  {author} {\bibinfo {author} {\bibfnamefont {M.}~\bibnamefont
  {Anđelković}}, \bibinfo {author} {\bibfnamefont {S.}~\bibnamefont
  {Milovanovic}}, \bibinfo {author} {\bibfnamefont {L.}~\bibnamefont {Covaci}},
   \emph {et~al.},\ }\href {\doibase 10.1021/acs.nanolett.9b04058} {\bibfield
  {journal} {\bibinfo  {journal} {Nano Lett.}\ }\textbf {\bibinfo {volume}
  {20}},\ \bibinfo {pages} {979} (\bibinfo {year} {2020})}\BibitemShut
  {NoStop}%
\bibitem [{\citenamefont {Wang}\ \emph
  {et~al.}(2019{\natexlab{a}})\citenamefont {Wang}, \citenamefont {Zihlmann},
  \citenamefont {Liu} \emph {et~al.}}]{wanglujun2019}%
  \BibitemOpen
  \bibfield  {author} {\bibinfo {author} {\bibfnamefont {L.}~\bibnamefont
  {Wang}}, \bibinfo {author} {\bibfnamefont {S.}~\bibnamefont {Zihlmann}},
  \bibinfo {author} {\bibfnamefont {M.-H.}\ \bibnamefont {Liu}},  \emph
  {et~al.},\ }\href {\doibase 10.1021/acs.nanolett.8b05061} {\bibfield
  {journal} {\bibinfo  {journal} {Nano. Lett.}\ }\textbf {\bibinfo {volume}
  {19}},\ \bibinfo {pages} {2371} (\bibinfo {year}
  {2019}{\natexlab{a}})}\BibitemShut {NoStop}%
\bibitem [{\citenamefont {Wang}\ \emph
  {et~al.}(2019{\natexlab{b}})\citenamefont {Wang}, \citenamefont {Yin},
  \citenamefont {Tóvári} \emph {et~al.}}]{wangyi2019}%
  \BibitemOpen
  \bibfield  {author} {\bibinfo {author} {\bibfnamefont {Y.}~\bibnamefont
  {Wang}}, \bibinfo {author} {\bibfnamefont {J.}~\bibnamefont {Yin}}, \bibinfo
  {author} {\bibfnamefont {E.}~\bibnamefont {Tóvári}},  \emph {et~al.},\
  }\href {\doibase 10.1126/sciadv.aay8897} {\bibfield  {journal} {\bibinfo
  {journal} {Sci. Adv.}\ }\textbf {\bibinfo {volume} {5}},\ \bibinfo {pages}
  {eaay8897} (\bibinfo {year} {2019}{\natexlab{b}})}\BibitemShut {NoStop}%
\bibitem [{\citenamefont {Khalaf}\ \emph {et~al.}(2019)\citenamefont {Khalaf},
  \citenamefont {Kruchkov}, \citenamefont {Tarnopolsky} \emph
  {et~al.}}]{khalaf2019}%
  \BibitemOpen
  \bibfield  {author} {\bibinfo {author} {\bibfnamefont {E.}~\bibnamefont
  {Khalaf}}, \bibinfo {author} {\bibfnamefont {A.~J.}\ \bibnamefont
  {Kruchkov}}, \bibinfo {author} {\bibfnamefont {G.}~\bibnamefont
  {Tarnopolsky}},  \emph {et~al.},\ }\href {\doibase
  10.1103/PhysRevB.100.085109} {\bibfield  {journal} {\bibinfo  {journal}
  {Phys. Rev. B}\ }\textbf {\bibinfo {volume} {100}},\ \bibinfo {pages}
  {085109} (\bibinfo {year} {2019})}\BibitemShut {NoStop}%
\bibitem [{\citenamefont {Ma}\ \emph {et~al.}(2022{\natexlab{b}})\citenamefont
  {Ma}, \citenamefont {Li}, \citenamefont {Xiao} \emph {et~al.}}]{double2023}%
  \BibitemOpen
  \bibfield  {author} {\bibinfo {author} {\bibfnamefont {Z.}~\bibnamefont
  {Ma}}, \bibinfo {author} {\bibfnamefont {S.}~\bibnamefont {Li}}, \bibinfo
  {author} {\bibfnamefont {M.-M.}\ \bibnamefont {Xiao}},  \emph {et~al.},\
  }\href {\doibase 10.1007/s11467-022-1220-z} {\bibfield  {journal} {\bibinfo
  {journal} {Front. Phys.}\ }\textbf {\bibinfo {volume} {18}},\ \bibinfo
  {pages} {13307} (\bibinfo {year} {2022}{\natexlab{b}})}\BibitemShut {NoStop}%
\bibitem [{\citenamefont {{Ding}}\ \emph {et~al.}(2022)\citenamefont {{Ding}},
  \citenamefont {{Liang}}, \citenamefont {{Ma}} \emph {et~al.}}]{ding2023}%
  \BibitemOpen
  \bibfield  {author} {\bibinfo {author} {\bibfnamefont {S.-P.}\ \bibnamefont
  {{Ding}}}, \bibinfo {author} {\bibfnamefont {M.}~\bibnamefont {{Liang}}},
  \bibinfo {author} {\bibfnamefont {Z.}~\bibnamefont {{Ma}}},  \emph {et~al.},\
  }\href {\doibase 10.48550/arXiv.2212.13393} {\ ,\ \bibinfo {pages}
  {arXiv:2212.13393} (\bibinfo {year} {2022})}\BibitemShut {NoStop}%
\bibitem [{\citenamefont {Park}\ \emph {et~al.}(2022)\citenamefont {Park},
  \citenamefont {Cao}, \citenamefont {Xia} \emph {et~al.}}]{Park20224}%
  \BibitemOpen
  \bibfield  {author} {\bibinfo {author} {\bibfnamefont {J.~M.}\ \bibnamefont
  {Park}}, \bibinfo {author} {\bibfnamefont {Y.}~\bibnamefont {Cao}}, \bibinfo
  {author} {\bibfnamefont {L.-Q.}\ \bibnamefont {Xia}},  \emph {et~al.},\
  }\href {\doibase 10.1038/s41563-022-01287-1} {\bibfield  {journal} {\bibinfo
  {journal} {Nat. Mater.}\ }\textbf {\bibinfo {volume} {21}},\ \bibinfo {pages}
  {877} (\bibinfo {year} {2022})}\BibitemShut {NoStop}%
\bibitem [{\citenamefont {Oka}\ and\ \citenamefont
  {Koshino}(2021)}]{koshino2021}%
  \BibitemOpen
  \bibfield  {author} {\bibinfo {author} {\bibfnamefont {H.}~\bibnamefont
  {Oka}}\ and\ \bibinfo {author} {\bibfnamefont {M.}~\bibnamefont {Koshino}},\
  }\href {\doibase 10.1103/PhysRevB.104.035306} {\bibfield  {journal} {\bibinfo
   {journal} {Phys. Rev. B}\ }\textbf {\bibinfo {volume} {104}},\ \bibinfo
  {pages} {035306} (\bibinfo {year} {2021})}\BibitemShut {NoStop}%
\bibitem [{\citenamefont {Mao}\ and\ \citenamefont
  {Senthil}(2021)}]{Maodan2021}%
  \BibitemOpen
  \bibfield  {author} {\bibinfo {author} {\bibfnamefont {D.}~\bibnamefont
  {Mao}}\ and\ \bibinfo {author} {\bibfnamefont {T.}~\bibnamefont {Senthil}},\
  }\href {\doibase 10.1103/PhysRevB.103.115110} {\bibfield  {journal} {\bibinfo
   {journal} {Phys. Rev. B}\ }\textbf {\bibinfo {volume} {103}},\ \bibinfo
  {pages} {115110} (\bibinfo {year} {2021})}\BibitemShut {NoStop}%
\bibitem [{\citenamefont {Long}\ \emph {et~al.}(2023)\citenamefont {Long},
  \citenamefont {Zhan}, \citenamefont {Pantale\'on} \emph
  {et~al.}}]{yuanprb2023}%
  \BibitemOpen
  \bibfield  {author} {\bibinfo {author} {\bibfnamefont {M.}~\bibnamefont
  {Long}}, \bibinfo {author} {\bibfnamefont {Z.}~\bibnamefont {Zhan}}, \bibinfo
  {author} {\bibfnamefont {P.~A.}\ \bibnamefont {Pantale\'on}},  \emph
  {et~al.},\ }\href {\doibase 10.1103/PhysRevB.107.115140} {\bibfield
  {journal} {\bibinfo  {journal} {Phys. Rev. B}\ }\textbf {\bibinfo {volume}
  {107}},\ \bibinfo {pages} {115140} (\bibinfo {year} {2023})}\BibitemShut
  {NoStop}%
\bibitem [{\citenamefont {Cea}\ \emph {et~al.}(2020)\citenamefont {Cea},
  \citenamefont {Pantale\'on},\ and\ \citenamefont {Guinea}}]{guineaprb2020}%
  \BibitemOpen
  \bibfield  {author} {\bibinfo {author} {\bibfnamefont {T.}~\bibnamefont
  {Cea}}, \bibinfo {author} {\bibfnamefont {P.~A.}\ \bibnamefont
  {Pantale\'on}}, \ and\ \bibinfo {author} {\bibfnamefont {F.}~\bibnamefont
  {Guinea}},\ }\href {\doibase 10.1103/PhysRevB.102.155136} {\bibfield
  {journal} {\bibinfo  {journal} {Phys. Rev. B}\ }\textbf {\bibinfo {volume}
  {102}},\ \bibinfo {pages} {155136} (\bibinfo {year} {2020})}\BibitemShut
  {NoStop}%
\bibitem [{\citenamefont {Shi}\ \emph {et~al.}(2021)\citenamefont {Shi},
  \citenamefont {Zhu},\ and\ \citenamefont {MacDonald}}]{shijingtian2021}%
  \BibitemOpen
  \bibfield  {author} {\bibinfo {author} {\bibfnamefont {J.}~\bibnamefont
  {Shi}}, \bibinfo {author} {\bibfnamefont {J.}~\bibnamefont {Zhu}}, \ and\
  \bibinfo {author} {\bibfnamefont {A.~H.}\ \bibnamefont {MacDonald}},\ }\href
  {\doibase 10.1103/PhysRevB.103.075122} {\bibfield  {journal} {\bibinfo
  {journal} {Phys. Rev. B}\ }\textbf {\bibinfo {volume} {103}},\ \bibinfo
  {pages} {075122} (\bibinfo {year} {2021})}\BibitemShut {NoStop}%
\bibitem [{\citenamefont {Jiang}\ and\ \citenamefont
  {Zhang}(2014)}]{zhangpingwen2014}%
  \BibitemOpen
  \bibfield  {author} {\bibinfo {author} {\bibfnamefont {K.}~\bibnamefont
  {Jiang}}\ and\ \bibinfo {author} {\bibfnamefont {P.}~\bibnamefont {Zhang}},\
  }\href {\doibase https://doi.org/10.1016/j.jcp.2013.08.034} {\bibfield
  {journal} {\bibinfo  {journal} {J. Comput. Phys.}\ }\textbf {\bibinfo
  {volume} {256}},\ \bibinfo {pages} {428} (\bibinfo {year}
  {2014})}\BibitemShut {NoStop}%
\bibitem [{\citenamefont {Cances}\ \emph {et~al.}(2017)\citenamefont {Cances},
  \citenamefont {Cazeaux},\ and\ \citenamefont {Luskin}}]{luskin2017}%
  \BibitemOpen
  \bibfield  {author} {\bibinfo {author} {\bibfnamefont {E.}~\bibnamefont
  {Cances}}, \bibinfo {author} {\bibfnamefont {P.}~\bibnamefont {Cazeaux}}, \
  and\ \bibinfo {author} {\bibfnamefont {M.}~\bibnamefont {Luskin}},\ }\href
  {\doibase 10.1063/1.4984041} {\bibfield  {journal} {\bibinfo  {journal} {J.
  Comput. Phys.}\ }\textbf {\bibinfo {volume} {58}},\ \bibinfo {pages} {063502}
  (\bibinfo {year} {2017})}\BibitemShut {NoStop}%
\bibitem [{\citenamefont {Massatt}\ \emph {et~al.}(2018)\citenamefont
  {Massatt}, \citenamefont {Carr}, \citenamefont {Luskin} \emph
  {et~al.}}]{massatt2018}%
  \BibitemOpen
  \bibfield  {author} {\bibinfo {author} {\bibfnamefont {D.}~\bibnamefont
  {Massatt}}, \bibinfo {author} {\bibfnamefont {S.}~\bibnamefont {Carr}},
  \bibinfo {author} {\bibfnamefont {M.}~\bibnamefont {Luskin}},  \emph
  {et~al.},\ }\href {\doibase 10.1137/17M1141035} {\bibfield  {journal}
  {\bibinfo  {journal} {Multiscale Model. Simul.}\ }\textbf {\bibinfo {volume}
  {16}},\ \bibinfo {pages} {429–451} (\bibinfo {year} {2018})}\BibitemShut
  {NoStop}%
\bibitem [{\citenamefont {{Zhou}}\ \emph {et~al.}(2019)\citenamefont {{Zhou}},
  \citenamefont {{Chen}},\ and\ \citenamefont {{Zhou}}}]{planewave2019}%
  \BibitemOpen
  \bibfield  {author} {\bibinfo {author} {\bibfnamefont {Y.}~\bibnamefont
  {{Zhou}}}, \bibinfo {author} {\bibfnamefont {H.}~\bibnamefont {{Chen}}}, \
  and\ \bibinfo {author} {\bibfnamefont {A.}~\bibnamefont {{Zhou}}},\ }\href
  {\doibase 10.1016/j.jcp.2019.02.003} {\bibfield  {journal} {\bibinfo
  {journal} {J. Comput. Phys.}\ }\textbf {\bibinfo {volume} {384}},\ \bibinfo
  {pages} {99} (\bibinfo {year} {2019})}\BibitemShut {NoStop}%
\bibitem [{\citenamefont {Ashcroft}\ and\ \citenamefont
  {Mermin}(1976)}]{solidstatebook}%
  \BibitemOpen
  \bibfield  {author} {\bibinfo {author} {\bibfnamefont {N.~W.}\ \bibnamefont
  {Ashcroft}}\ and\ \bibinfo {author} {\bibfnamefont {N.~D.}\ \bibnamefont
  {Mermin}},\ }\href@noop {} {\emph {\bibinfo {title} {Solid State Physics}}}\
  (\bibinfo  {publisher} {Cengage Learning},\ \bibinfo {year}
  {1976})\BibitemShut {NoStop}%
\bibitem [{Note1()}]{Note1}%
  \BibitemOpen
  \bibinfo {note} {See Supplemental Material at [URL] for the interpretation
  about the meaning of $n_c$.}\BibitemShut {Stop}%
\bibitem [{Note2()}]{Note2}%
  \BibitemOpen
  \bibinfo {note} {See Supplemental Material at [URL] for the proof of this
  statment.}\BibitemShut {Stop}%
\bibitem [{Note3()}]{Note3}%
  \BibitemOpen
  \bibinfo {note} {See Supplemental Material at [URL] for the calculation
  results with $G_2$ as PBZ.}\BibitemShut {Stop}%
\bibitem [{Note4()}]{Note4}%
  \BibitemOpen
  \bibinfo {note} {See Supplemental Material at [URL] for the example of
  commensurate case.}\BibitemShut {Stop}%
\bibitem [{\citenamefont {Casati}\ \emph {et~al.}(1989)\citenamefont {Casati},
  \citenamefont {Guarneri},\ and\ \citenamefont {Shepelyansky}}]{TIP1989}%
  \BibitemOpen
  \bibfield  {author} {\bibinfo {author} {\bibfnamefont {G.}~\bibnamefont
  {Casati}}, \bibinfo {author} {\bibfnamefont {I.}~\bibnamefont {Guarneri}}, \
  and\ \bibinfo {author} {\bibfnamefont {D.~L.}\ \bibnamefont {Shepelyansky}},\
  }\href {\doibase 10.1103/PhysRevLett.62.345} {\bibfield  {journal} {\bibinfo
  {journal} {Phys. Rev. Lett.}\ }\textbf {\bibinfo {volume} {62}},\ \bibinfo
  {pages} {345} (\bibinfo {year} {1989})}\BibitemShut {NoStop}%
\bibitem [{Note5()}]{Note5}%
  \BibitemOpen
  \bibinfo {note} {See Supplemental Material at [URL] for the details about
  momentum edge states.}\BibitemShut {Stop}%
\bibitem [{\citenamefont {Chen}\ \emph
  {et~al.}(2021{\natexlab{b}})\citenamefont {Chen}, \citenamefont {Zhou},\ and\
  \citenamefont {Zhou}}]{localization2021}%
  \BibitemOpen
  \bibfield  {author} {\bibinfo {author} {\bibfnamefont {H.}~\bibnamefont
  {Chen}}, \bibinfo {author} {\bibfnamefont {A.}~\bibnamefont {Zhou}}, \ and\
  \bibinfo {author} {\bibfnamefont {Y.}~\bibnamefont {Zhou}},\ }\href {\doibase
  https://doi.org/10.1016/j.commatsci.2020.110242} {\bibfield  {journal}
  {\bibinfo  {journal} {Comput. Mater. Sci.}\ }\textbf {\bibinfo {volume}
  {188}},\ \bibinfo {pages} {110242} (\bibinfo {year}
  {2021}{\natexlab{b}})}\BibitemShut {NoStop}%
\bibitem [{\citenamefont {Lang}\ \emph {et~al.}(2012)\citenamefont {Lang},
  \citenamefont {Cai},\ and\ \citenamefont {Chen}}]{caixiaoming2011}%
  \BibitemOpen
  \bibfield  {author} {\bibinfo {author} {\bibfnamefont {L.-J.}\ \bibnamefont
  {Lang}}, \bibinfo {author} {\bibfnamefont {X.}~\bibnamefont {Cai}}, \ and\
  \bibinfo {author} {\bibfnamefont {S.}~\bibnamefont {Chen}},\ }\href {\doibase
  10.1103/PhysRevLett.108.220401} {\bibfield  {journal} {\bibinfo  {journal}
  {Phys. Rev. Lett.}\ }\textbf {\bibinfo {volume} {108}},\ \bibinfo {pages}
  {220401} (\bibinfo {year} {2012})}\BibitemShut {NoStop}%
\bibitem [{Note6()}]{Note6}%
  \BibitemOpen
  \bibinfo {note} {See Supplemental Material at [URL] for the relation to the
  experimental model.}\BibitemShut {Stop}%
\bibitem [{Note7()}]{Note7}%
  \BibitemOpen
  \bibinfo {note} {See Supplemental Material at [URL] for the proof of this
  statement.}\BibitemShut {Stop}%
\bibitem [{Note8()}]{Note8}%
  \BibitemOpen
  \bibinfo {note} {See Supplemental Material at [URL] for the calculation
  details.}\BibitemShut {Stop}%
\end{thebibliography}%


\clearpage
\widetext

\newcommand{\bk}{\bm{k}}
\newcommand{\bq}{\bm{q}}
\newcommand{\btk}{\widetilde{\bm{k}}}
\newcommand{\btq}{\widetilde{\bm{q}}}
\newcommand{\br}{\bm{r}}
\newcommand{\cop}{\hat{c}}
\newcommand{\dop}{\hat{d}}
\newcommand{\xmark}{\ding{55}}
\def\Red#1{\textcolor{red}{#1}}
\def\Blue#1{\textcolor{blue}{#1}}

\begin{center}
\textbf{\large Supplementary Materials for: Energy Spectrum Theory of  Incommensurate Systems}
\end{center}

\maketitle

\setcounter{equation}{0}
\setcounter{figure}{0}
\setcounter{table}{0}
\setcounter{page}{1}
\makeatletter
\renewcommand{\theequation}{S\arabic{equation}}
\renewcommand{\thefigure}{S\arabic{figure}}
\renewcommand{\bibnumfmt}[1]{[S#1]}


\section{I. The equivalent momenta in PBZ}
For the BIP model, the incommensurate central equations indicate that all the momenta in the set $Q_{q}=\left\{ k | k=q+m G_1+nG_2 : m,n\in \mathbb{Z} \right\}$ are all equivalent. The equivalent relations  imply a fact: for a given $q$, there are $N_E$ equivalent momenta in the PBZ, which do not overlap with each other in the incommensurate case. Here, we give a proof of this statement. 

$N_E$ is the total number of the all the allowed $n$, and  each $n$ corresponds to an equivalent moment $(q+nG_2)+mG_1$ in PBZ, where $m$ is an integer. 
Suppose that $q_n$ and $q_{n'}$ are two equivalent momenta in PBZ with $n \neq n'$, i.e. 
 \begin{equation}\label{Set}
 \begin{aligned}
    q_{n} &= (q +n G_2) + m_{n}G_1    \\
    q_{n'} &= (q + n' G_2)+ m_{n'}G_1  
  \end{aligned}
  \end{equation}
where $n$, $n'$, $m_n$ ane $m_{n'}$ are integers. If the two momenta coincide, we get
\begin{equation}
    q +  nG_2 + m_{n}G_1  = q  + n'G_2 + m_{n'}G_1
\end{equation}
which means
\begin{equation}
    \frac{G_1}{G_2} = \frac{m_{n'}-m_{n}}{n - n'} = \frac{1}{\alpha}.
\end{equation}
$\alpha$ now becomes a rational number, which is clearly in contrast to the incommensurate assumption. So, the conclusion is that all the $N_E$ equivalent momenta in PBZ will never overlap with one another.  

This statement can be generalized to the 2D incommensurate systems. Here, we use the 2D moire quasicrystal as an example. 
For a given $\mathbf{q}$,  all the equivalent momenta can be expressed as $\mathbf{q}+m_1 \mathbf{G_a}+m_2 \mathbf{G_b} + n_1 \mathbf{\Tilde{G}_a} + n_2 \mathbf{\Tilde{G}_b}$. Suppose that $\mathbf{q_n}$ and $\mathbf{q_{n'}}$ are two equivalent momenta in PBZ with $(n_1,n_2) \neq (n'_1,n'_2)$. If they coincide, we have 
\begin{equation}
    \mathbf{q}+m_1 \mathbf{G_a}+m_2 \mathbf{G_b} + n_1 \mathbf{\Tilde{G}_a} + n_2 \mathbf{\Tilde{G}_b} =  \mathbf{q}+m'_1 \mathbf{G_a}+m'_2 \mathbf{G_b} + n'_1 \mathbf{\Tilde{G}_a} + n'_2 \mathbf{\Tilde{G}_b}
\end{equation}
It means 
\begin{equation}
    (m_1 -m'_1)\mathbf{G_a}+(m_2-m'_2) \mathbf{G_b} =  (n'_1-n_1) \mathbf{\Tilde{G}_a} + (n'_2-n_2) \mathbf{\Tilde{G}_b}
\end{equation}
where right side (left) side is a lattice vector of the first (second) periodic potential $V_1$ ($V_2$), i.e.~$ (m_1 -m'_1)\mathbf{G_a}+(m_2-m'_2) \mathbf{G_b}$ is a common lattice vector for both $V_1$ and $V_2$. Note that both $V_1$ and $V_2$ are square periodic potentials, which has $C_4$ symmetry. Therefore, we know that $ (m_1 -m'_1)R(\pi/4)\mathbf{G_a}+(m_2-m'_2) R(\pi/4)\mathbf{G_b}$ is a common lattice vector for both $V_1$ and $V_2$ as well. Such two common lattice vectors in momentum space actually corresponds to a supercell of the moire quasicrystal, which is in contrast to the incommensurate assumption. So, we get the conclusion that all the equivalent momenta will never overlap with each other. 

\section{II. Momentum edge states}
\begin{figure}[tbp!]
    \centering
    \includegraphics[width=0.6\textwidth]{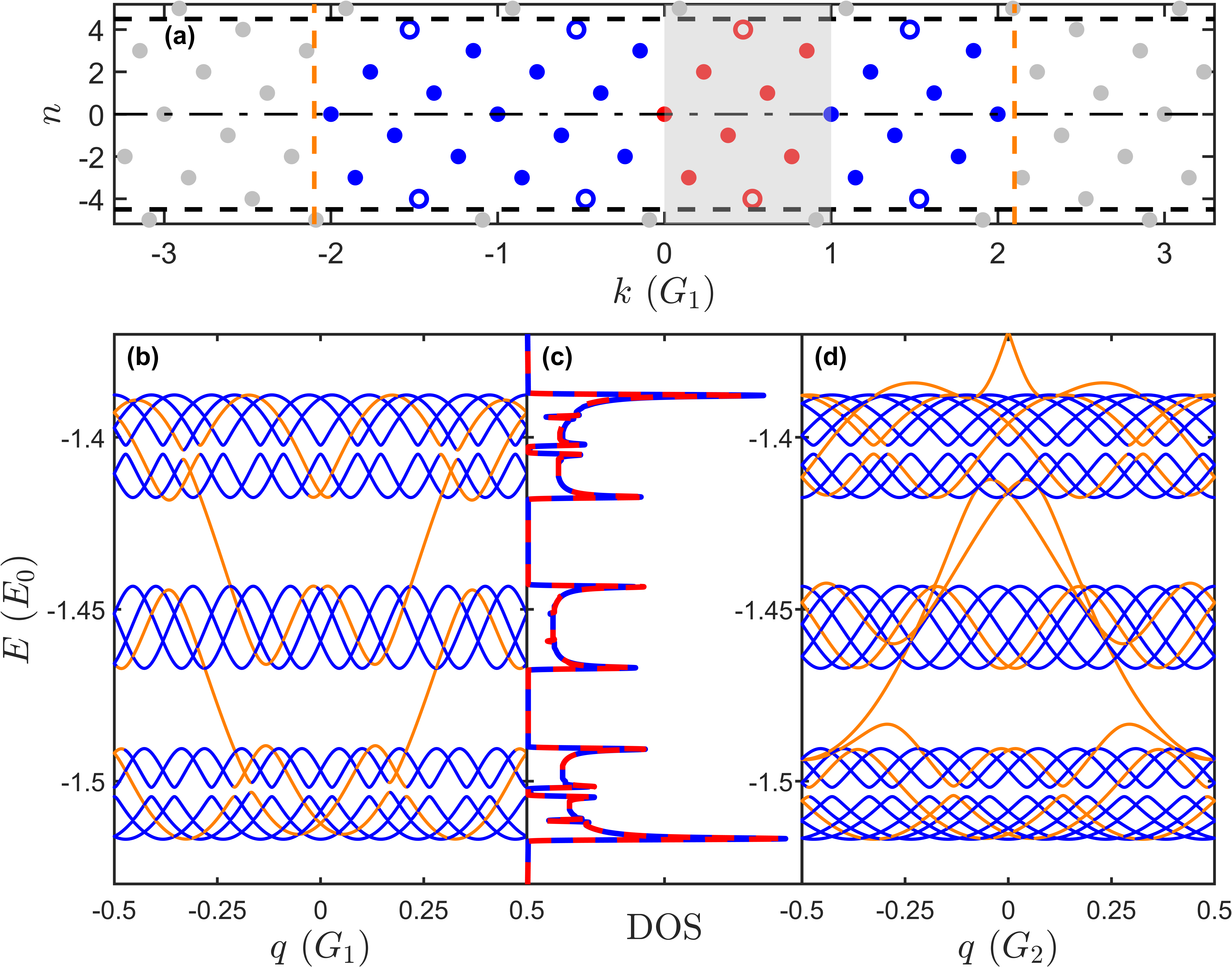}
    \caption{(a) Schematic of the momentum edge states in the BIP model with $\alpha=(\sqrt{5}-1)/2$.  The $n_c$ induced momentum edge states are represents as the hollow dots.
    (b) is same as the Fig.~1 (b) of the main text. Parameters:  $V_1=8E_0$, $V_2=0.06E_0$, $n_c=8$, $\phi=0$, $k_c=4G_1$. 
         (d) is the results with $G_2$ as the PBZ. Parameters: $V_1=8E_0$, $V_2=0.06E_0$, $n_c=10$, $\phi=0$, $k_c=4G_1$.
       (c) DOS, blue lines represent (b), and red lines represent (d).
    }
    \label{figs1}
\end{figure}

As explained in the main text, the truncation of the $n_c$ will give rise to momentum edge states mostly in the energy gaps, due to appearance of the open boundary in momentum space. In principle, the truncation of $k_c$ will also give rise to open boundary, but these boundary states only appear in the high energy range, since that $k_c$ should correspond to the maximum energy of the plane waves. Therefore, for the energy region that we are interested in, only the $n_c$ induced momentum edge states has to be considered. In other words, the momentum edge states in Fig.~1 (b) of the main text are mainly distributed around the $n_c$ induced boundaries, which are are illustrated  in Fig.~\ref{figs1} (a) (hollow dots).

Because that the momentum edge states depend on the truncation, they are artificially induced states, which should be eliminated in the calculations. In practice, we identify the momentum edge states by examining the wave function distribution in the momentum space. The eigenstates will be deleted from the final results, if they distribute mainly  around the boundaries. Note that, the boundary is not just a line, but has width. So, we define $n_e$ as the boundary width, which means the outer most $n_e$ sites are viewed as the boundary regions. The $N_E$ should be corrected when some equivalent momenta are near the boundary, i.e.~$N_E=2(n_c-n_e) + 1$.  

Of course, The criterion of the boundary states is not unique, which  can be further improved in subsequent works. However, because that the bulk states actually dominate in such systems, small changes of the boundary criteria will not affect the final results.  

\section{III. The physical significance  of the truncation}
To calculate the incommensurate energy spectrum, we need to introduce a truncation of $(m,n)$ to get a finite-dimensional Hamiltonian matrix $H(q)$. As discussed in the BIP model, the used truncation is $|n|<n_c$ and $|k|<k_c$, where $n_c$ and $k_c$ are two provided truncation constants. The physical meaning of $k_c$ is very clear, which represents the cutoff of the energy of plane waves. Meanwhile, $n_c$ also has a clear physical significance, which indeed reflects the minimal interval between the equivalent momenta  for given $q$.  Actually, due to the incommensurability, the combination of $G_1$ and $G_2$, i.e.~$mG_1 + nG_2$, can give rise to arbitrarily small momentum, which is essentially different from the commensurate case.  When $|n|<n_c$, no matter what the value of $m$ is,  $|mG_1 + nG_2|$ always has a minimum value, which depends on the value of $n_c$. Such physical pictures of the truncation is valid  for all the incommensurate systems. 

\section{IV. Primary Brillouin Zone}
In principle, it is better to use a deep potential as the PBZ, and a shallow potential as a perturbation. The advantage lies in the fact that shallow perturbation can lead to a small truncation $n_c$, as demonstrated in the main text.  The converse is also valid, but the truncation should be larger.
In Fig.~\ref{figs1} (c) and (d), we use the interval $[0,G_2)$ as the PBZ and calculate the DOS and energy spectrum. We see that, with a larger  truncation $n_c$, we get the same DOS. 

\section{V. The BIP model}
\begin{figure}[tbp!]
    \centering
    \includegraphics[width=0.6\textwidth]{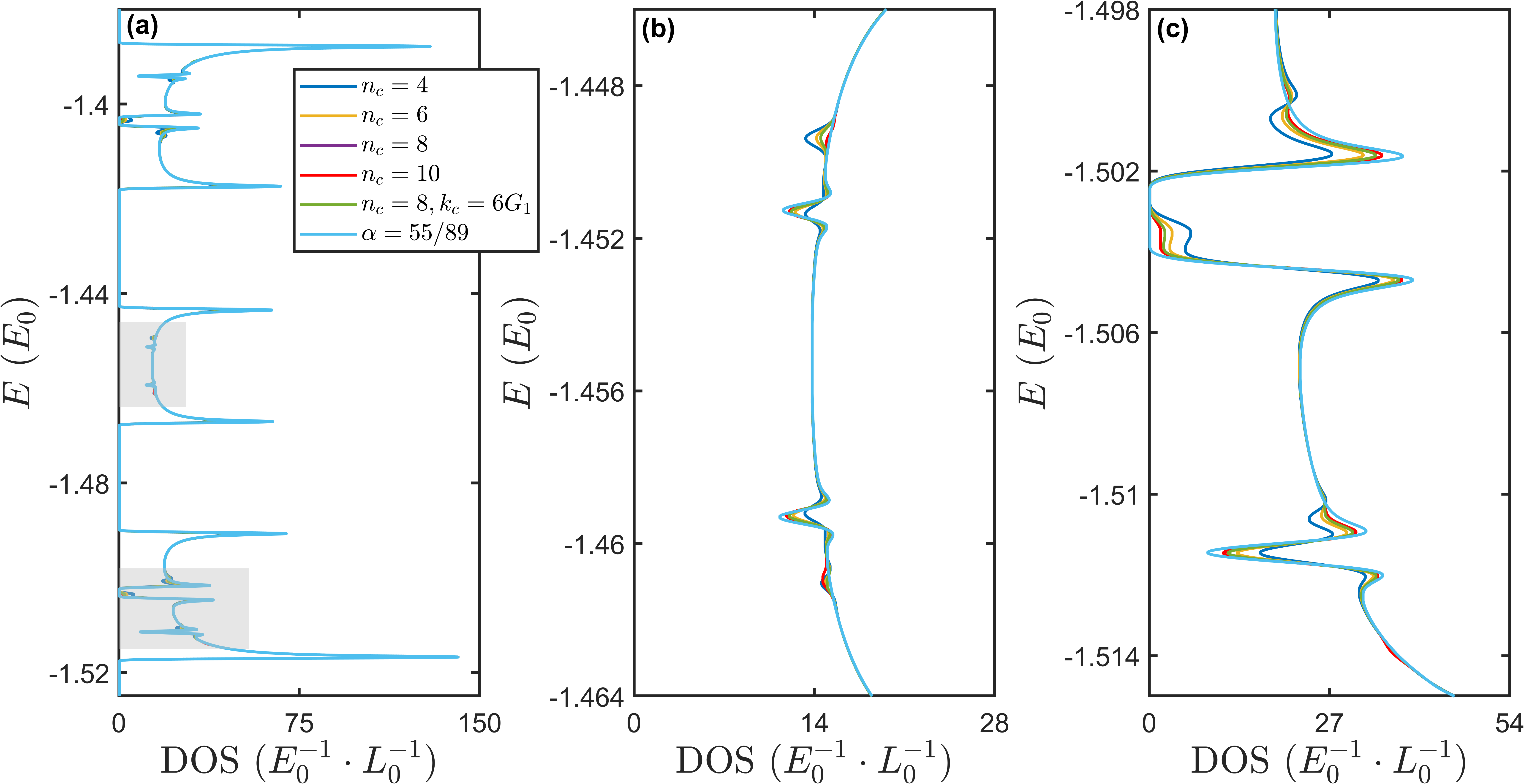}
    \caption{ The DOS of the BIP model with different $n_c$ and $k_c$. (b) and (c) are the enlarged DOS plots for the gray regions in (a). Parameters:
    $V_1=8E_0,V_2=0.06E_0,\alpha=(\sqrt{5}-1)/2,k_c=4G_1$.
    }
    \label{figs2}
\end{figure}

Here, we give some calculation details of the BIP model. Fig.~\ref{figs2} plots the calculated DOS with different truncation $n_c$ and $k_c$, which illustrates the convergence of the results.  The DOS results show that we can get a convergent numerical results even with truncation $n_c=4$. And the results from the IES theory are in good agreement with the commensurate approximation as well. To calculate the DOS, we uniformly choose 1000 $q$ points in the PBZ. For each $q$ point, the $H(q)$ is a $137 \times 137$ matrix with $n_c=8$ and $k_c=4G_1$.

\section{VI. The TIP model}
\begin{figure}[tbp!]
    \centering
    \includegraphics[width=0.6\textwidth]{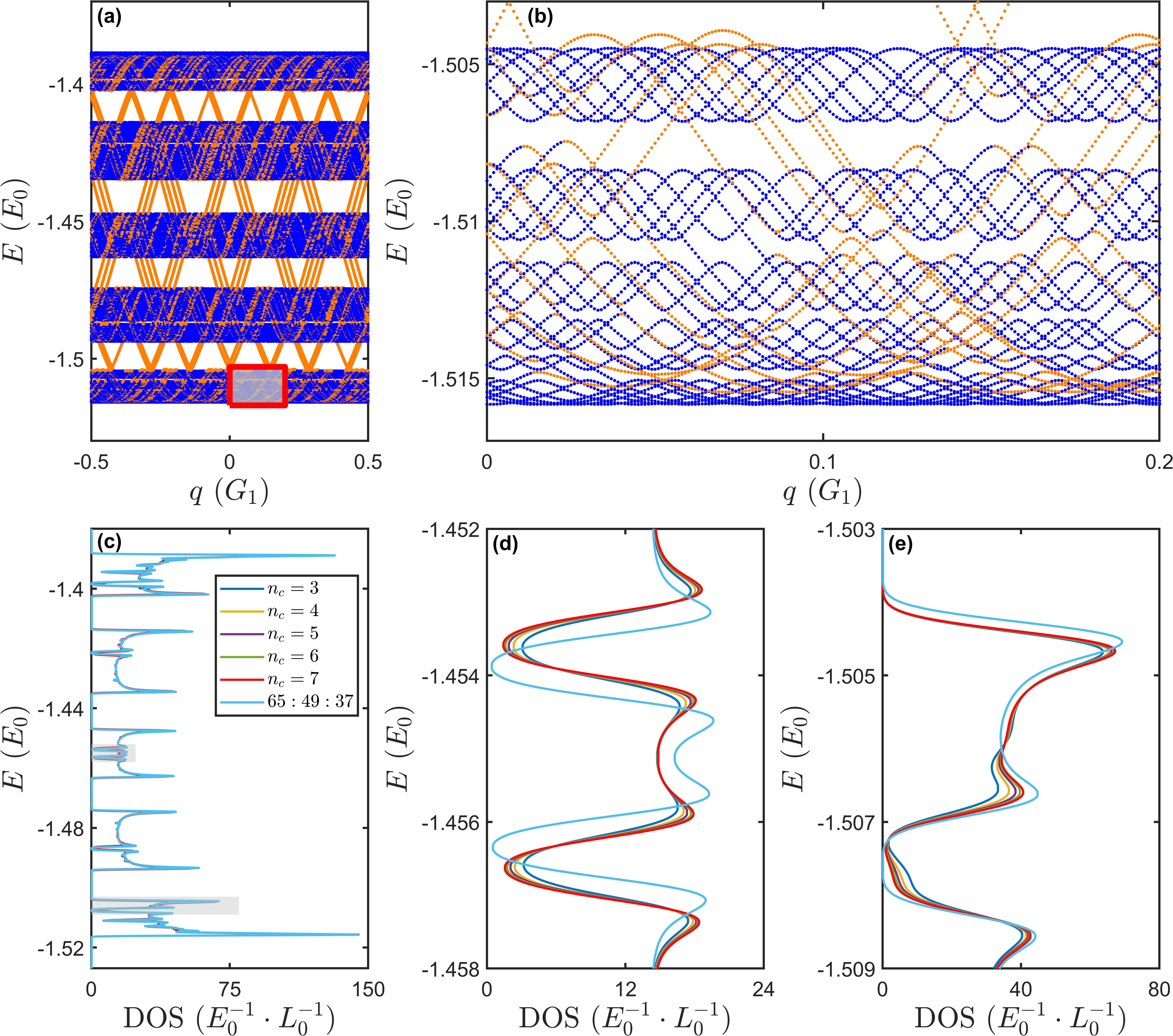}
    \caption{ The TIP model. (a) is the energy spectrum diagram, the same as the Fig.~2 of the main text. (b) and (c) are the  enlarged energy spectrum plot for the region of the red box in  (a).
    (c) is the DOS calculated with different truncation $n_c$, and the results of a commensurate approximation is also given. (d) and (e) are enlarged DOS plots for the gray regions in (c).     Parameters:
    $V_1=8E_0,V_2=V_3=0.03E_0,k_c=4G_1$. The two irrational numbers are $\alpha_2=\lambda^{-1}$, $\alpha_3=\lambda^{-2}$, where $\lambda=1.3247 \cdots$ is the root of the equation $x^3-x-1=0$~\cite{TIP1989}.
    }
    \label{figs4}
\end{figure}

Here, we give some calculation details of the TIP model. Fig.~\ref{figs4} (a) is the same as the Fig.~2 of the main text. Then,  we plot the enlarged energy spectrum diagram (region in the red box) in Fig.~\ref{figs4} (b). The DOS with various truncations are shown in Fig.~\ref{figs4} (c) to illustrate the numerical convergence, where Fig.~\ref{figs4} (c) and (d) are the enlarged DOS plots in the gray regions. The DOS of a commensurate approximation ($G_1:G_2:G_3 \approx 65:49:37$) is also given as a comparison . First, we see that a convergent DOS can be achieved even with $n_c=3$.  Meanwhile,  Fig.~\ref{figs4} (d) and (e) indicate that the IES theory may give better results than that of the common commensurate approximation. It is because that, with the IES theory,  the predicted position of peaks always remains unchanged with increasing $n_c$.  In Fig.~\ref{figs4},  $H(q)$  is a $393\times 393$ matrix when $n_c=3$.

\section{VII. Moire quasicrystal}
First, we would like to show that the moire quasicrystal model here is exactly the same as that in Ref.~23 of the main text. In Ref.~23, the potential is 
\begin{align}
     V(r) &=V_0 \sum_{i=1}^{D=4} \cos^2(\frac{\mathbf{G_i}}{2}\cdot r)   \\
          &= \frac{V_0}{2}\sum_{i=1}^{D=4} \cos(\mathbf{G_i}\cdot r)+2V_0
\end{align}
If we set  $\mathbf{G_1}=\mathbf{G_a}$, $\mathbf{G_3}=\mathbf{G_b}$, $\mathbf{G_2}=\mathbf{\Tilde{G}_a}$ and  $\mathbf{G_4}=\mathbf{\Tilde{G}_b}$, we then obtain the moire quasicrystal potential in the main text, except for a factor $\frac{1}{2}$ for $V_0$ and a constant potential shift.

\begin{figure}[tbp!]
    \centering
    \includegraphics[width=0.6\textwidth]{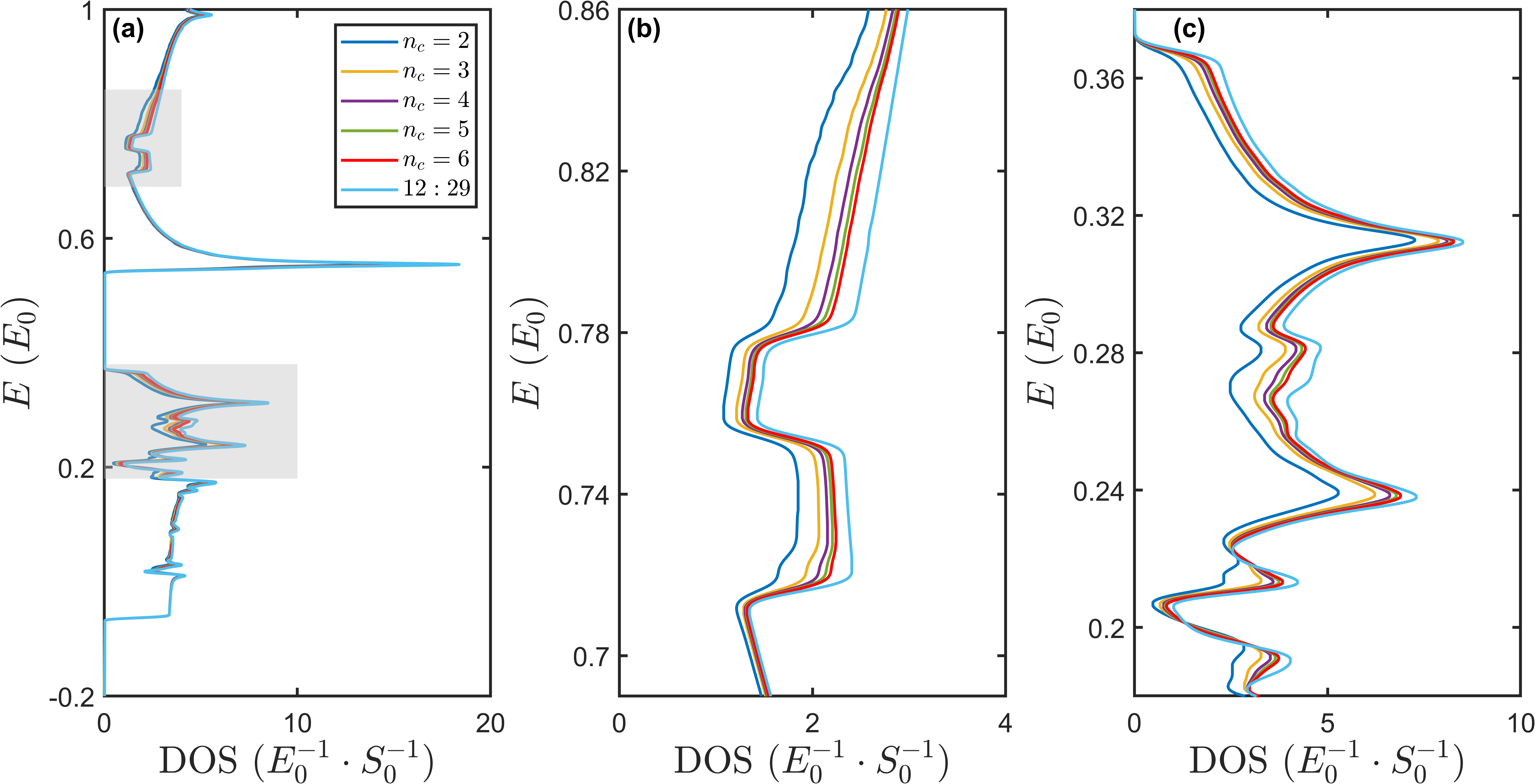}
    \caption{The DOS of moir\'{e} quasicrystal with different $n_c$. (b) and (c) are the enlarged plots for the gray regions in (a). 
    Parameters:
    $V_0=1E_0, k_c=2.1 |\mathbf{G_a}|$.    
    }
    \label{figs5}
\end{figure}

In Fig.~\ref{figs5}, we plot the calculated DOS with different truncation to illustrate the convergence.  The results of a commensurate approximation  structure is given as well. Here,  a commensurate approximation structure is described by two integers $(m,n)$ with
\begin{equation}
    \cos\theta_{mn}=\frac{n^2-m^2}{m^2+n^2},
\end{equation}
where $\theta_{mn}$ is the twisted angle~\cite{Ye2020}. For the moire quasicrystal, $(m,n)=(12,29)$ corresponds to $\theta_{mn} \approx 44.96^\circ$, which is a reasonable commensurate approximation. The calculated results indicate that the truncation $n_c=4$ has already given a satisfied DOS results. 
The DOS is calculated with a $400\times 400$ mesh in PBZ.

\section{VIII. The commensurate cases}
\begin{figure}[tbp!]
    \centering
    \includegraphics[width=0.6\textwidth]{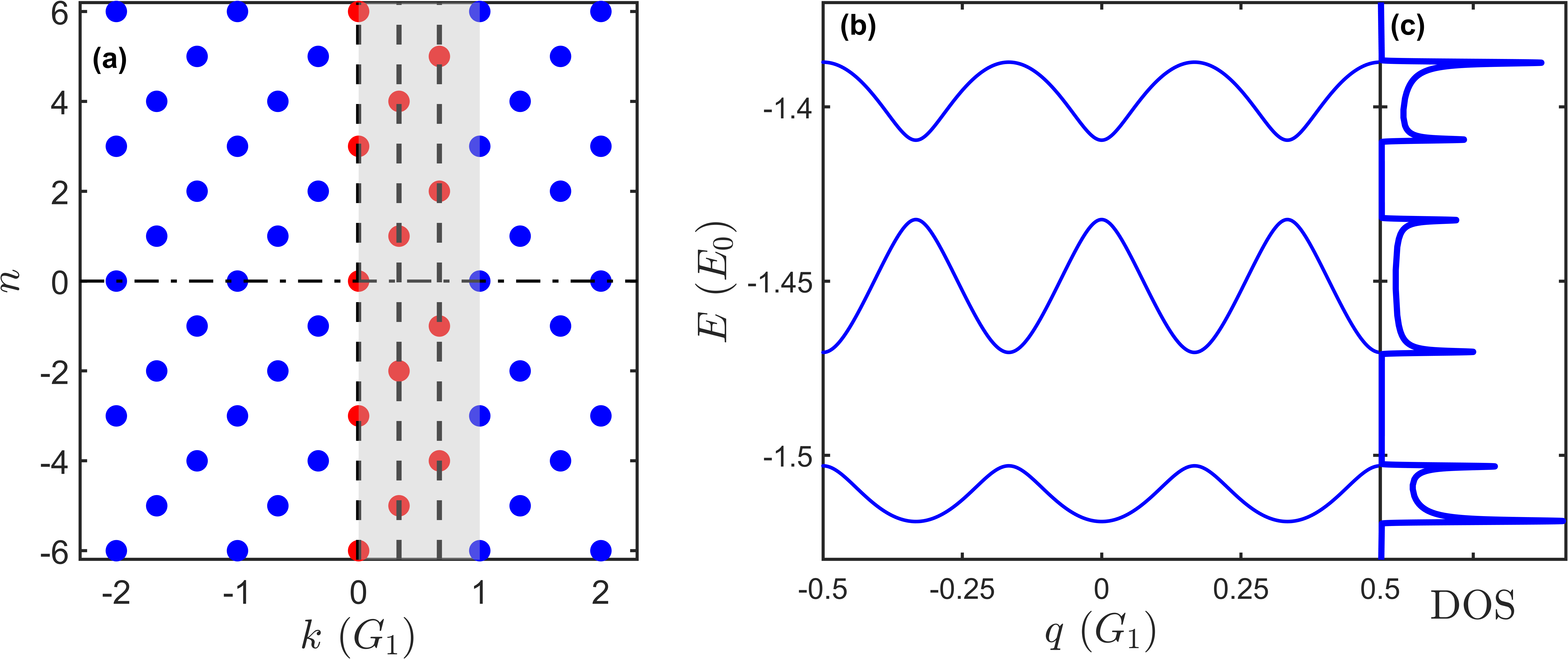}
    \caption{The commensurate bichromatic potential model with $G_1:G_2=3:1$. (a) shows the equivalent momenta. (b) and (c) are the energy spectrum and DOS. Parameters: $k_c=4G_1,V_1=8E_0,V_2=0.06E_0,\phi=0$. 
    }
    \label{figs3}
\end{figure}

Here, we use the bichromatic potential model as an example to illustrate that the IES formulae are also valid for the commensurate case. Specifically, we consider a commensurate situation: $G_1:G_2=3:1$. Based on the Eq.~(2) of the main text, we can also plot the equivalent momenta for a given $q$ with $G_1$ as the PBZ, see Fig.~\ref{figs3} (a). Clearly, distinct from the incommensurate case,   the equivalent momenta will overlap in such commensurate case. Thus, we eventually get three distinct equivalent momenta in the PBZ, i.e.~$N_E=3$,  which is irrelevant to the $n_c$. Note that $[0,G_1/3)$ is just the FBZ of such commensurate case. Therefore, Eq.~(4) and (5) of the main text do give the correct DOS and expectation values. The energy spectum and DOS are given in Fig.~\ref{figs3} (b). This example  clearly demonstrates that the IES theory actually is applicable to the commensurate cases. So, the IES formulae in fact provides a unified theoretical framework to treat the multi periodic potential models, no matter whether it is incommensurate or not. In practical calculations, the only issue is to correctly determine $N_E$ by counting the distinct equivalent momenta in the PBZ. Note that in the commensurate cases, $n_c$ has no effect and $k_c$ is the only meaningful truncation. 

\end{document}